\documentclass[10pt,journal,compsoc]{IEEEtran}
\ifCLASSOPTIONcompsoc
  \usepackage[nocompress]{cite}
\else
  \usepackage{cite}
\fi
\ifCLASSINFOpdf
\else
\fi
\usepackage{amsfonts}    
\usepackage{color}
\usepackage{graphicx}
\usepackage[dvips]{epsfig}
\usepackage{graphics}
\usepackage{cite}
\usepackage{arydshln}
\usepackage{amsmath} 
\usepackage{amssymb}  
\usepackage{subfigure}
\usepackage{subfloat}
\usepackage{caption}
\usepackage{float}

\def\bE{\mathbb{E}}

\def\cD{\mathcal{D}}
\def\cE{\mathcal{E}}

\def\cM{\mathcal{M}}

\def\cT{\mathcal{T}}

\def \qed {\hfill \vrule height6pt width 6pt depth 0pt}
\def\bee{\begin{equation}}
\def\ene{\end{equation}}
\def\been{\begin{equation*}}
\def\enen{\end{equation*}}
\def\beq{\begin{eqnarray}}
\def\enq{\end{eqnarray}}

\newtheorem{pro}{Proposition}[section]
\newtheorem{lem}{Lemma}[section]
\newtheorem{rem}{Remark}[section]
\newtheorem{cor}{Corollary}[section]
\newtheorem{thm}{Theorem}[section]

\begin{document}
%
\title{Economic Analysis of Rollover and Shared Data Plans}
%
%
%
%

\author{Xuehe~Wang,~\IEEEmembership{Member,~IEEE,}
        and~Lingjie~Duan,~\IEEEmembership{Senior~Member,~IEEE}
\IEEEcompsocitemizethanks{\IEEEcompsocthanksitem This work was supported by the Singapore Ministry of Education Academic Research Fund Tier 2 under Grant MOE2016-T2-1-173. X. Wang and L. Duan are with the Pillar of Engineering Systems and Design, Singapore University of Technology and Design, Singapore 487372. E-mail: xuehe\_wang@sutd.edu.sg; lingjie\_duan@sutd.edu.sg.
}
}

\IEEEtitleabstractindextext{%
\begin{abstract}

In today's growing data market, wireless service providers (WSPs) compete severely to attract users by announcing innovative data plans. Two of the most popular innovative data plans are rollover and shared data plans, where the former plan allows a user to keep his unused data quota to next month and the latter plan allows users in a family to share unused data. As a pioneer to provide such data plans, a WSP faces immediate revenue loss from existing users who pay less overage charges due to less data over-usage, but his market share increases gradually by attracting new users and those under the other WSPs. In some countries, WSPs have asymmetric timing for providing such innovative data plans, while some other markets' WSPs have symmetric timing or no planning. This raises the question of why and when the competitive WSPs should offer the new data plans. This paper provides game theoretic modelling and analysis of the WSPs' timing of offering innovative data plans, by considering new user arrival and dynamic user churn between WSPs. Our equilibrium analysis shows that the WSP with small market share prefers to announce the innovative data plan first to attract more users, while the WSP with large market share prefers to announce later to avoid the immediate revenue loss. In a market with many new users, WSPs with similar market shares will offer the data plans simultaneously, but these WSPs facing few new users may not offer any new plan. Perhaps surprisingly, WSPs' profits can decrease with new user number and they may not benefit from the option of innovative data plans. Finally, unlike rollover data plan, we show that the timing of shared data plan further depends on the composition of users.

\end{abstract}

}

\maketitle

\IEEEdisplaynontitleabstractindextext

%
\IEEEpeerreviewmaketitle


%
%
%
%

%

\section{Introduction}

\IEEEPARstart{T}{he} development of mobile technology has been driving the rapid growth of mobile devices (e.g., smartphones and iPad) and the users' data consumption. To attract more users and increase market shares, competitive wireless service providers (WSPs) have announced new alluring data plans to save users' costs. Rollover data plan is one of the most popular innovative data plans, which allows the users to keep the current month's unused data quota to next month without any additional charge. Take the USA market as an example. AT\&T has freely upgraded its users' data plans to include rollover, where a user's left data from last month can be used after the current month's data quota is used up. Another popular innovative plan is shared data plan, which allows two or more users (devices) to share a common pool of data quota in each month. For example, China Mobile allows two users in the same city to combine their data plans without additional charge. Both rollover and shared data plans are attractive to (heavy) users by reducing their expected data costs in different ways: the former seeks inner-user data sharing by pooling an individual's monthly data quota over time, while the latter seeks inter-user data sharing by pooling two or more users' data quota in each month.

To a WSP's point of view, both rollover and shared data plans reduce its overage charges from heavy users, but can increase its market share by attracting new users and those from its WSP competitors. In reality, we observe that the WSPs' timing of announcing the innovative data plans is very different in different countries. For example, in the saturated data market of the USA, T-Mobile with small market share is a pioneer to announce its rollover data plan in 2014. Though its immediate revenue from existing users reduced, it successfully increased its market share by more than 1 million users in merely a quarter \cite{tmobile2015}. The major WSPs, AT$\&$T and Verizon in the same market unveiled their rollover data plans sequentially in 2015 and 2016, respectively. Different from the USA market, China's market is fast growing with many new users annually, and we observe that China Mobile and China Unicom offered the rollover data plan at the same time \cite{Chinarollover2015}. However, in some saturated markets such as Singapore, the major WSPs SingTel and Starhub have similar market shares and neither provides the rollover data plan. Similar situations happened to the shared data plan case. For example, in some countries (e.g., the UK), the WSPs with diverse market shares sequentially provided the shared data plans \cite{sharedUK2013}, while in some countries (e.g., China), WSPs offered the shared data plans almost at the same time \cite{ChinashareUnicom,ChinashareMobile}. This motivates us to ask the key question in this paper: Why and when should the competitive WSPs announce the innovative data plans?

There are some recent works discussing shared and rollover data plans from a user's or a single WSP's perspective (\cite{sen2012economics, jin2014smart, zheng2016understanding, wang2017pricing}). \cite{sen2012economics} models and numerically analyzes a user's choice between individual and shared data plan. In \cite{jin2014smart}, the authors study a monopoly WSP's decision on the adoption of shared data plan, by examining the cost for such a new service. \cite{zheng2016understanding} analyzes the benefits of rollover data for a monopoly WSP and its users. And the work in \cite{wang2017pricing} further designs the price of the rollover data plan to maximize the WSP's revenue. However, all these works only consider the monopoly WSP, and the competition between the WSPs (common in almost all data markets) has not been investigated yet. There are some prior works on generic network's competition or user dynamics (e.g., \cite{musacchio2006game}, \cite{duan2015economic}, \cite{duan2012duopoly}). The work in \cite{musacchio2006game} studies two interconnected service providers' timing of upgrading architecture, where one provider's upgrade also benefits the other due to the network effect. This ``free-rider" effect gives the other operator a temptation to postpone its upgrade or even not upgrade. The work in \cite{duan2015economic} studies the cellular operators' timing of 4G network upgrade and shows that operators select different upgrade times to avoid severe competition. However, these works focus on the upgrade competition between symmetric operators, where operators are assumed to have identical market share. \cite{duan2012duopoly} studies the duopoly competition in spectrum leasing market but only looked at a static model in one-shot. \cite{gong2017social} analyzes the user dynamics under innovative mobile social services rather than innovative data plans. In our paper, we look at a different problem without free riding effect and will show that the disparity of market shares will significantly affect the WSPs' timing of offering innovative data plans under users dynamics.

In this paper, we analyze the timing of competitive WSPs to announce the rollover or shared data plan, and answer why the WSPs in some countries prefer to offer simultaneously yet some other countries have asymmetric timing or no planning for innovative data plans. Our key novelty and main contributions are summarized as follows.
\begin{itemize}
  \item \emph{Economics of innovative data plans in competitive markets:} To our best knowledge, this is the first paper to provide game theoretic modelling and analysis of innovative data plans in competitive data markets. We practically model the dynamic user arrival and user churn between WSPs, and analytically characterize the pros and cons of innovative data plans for each WSP. We formulate the WSPs' interactions as a non-cooperative game, where each WSP seeks the trade-off between the increase of market share and the loss of overage charges.
  \item \emph{Equilibrium timing for innovative data plans:} By analyzing the non-cooperative timing game, our equilibrium analysis shows that the WSP with small market share would like to offer the rollover or shared data plan first to attract significantly more users, while the WSP with large market share prefers to announce later to avoid the immediate revenue (overage charge) loss. In a fast-growing market with many new user arrivals, we show that WSPs with similar market shares will offer the data plans simultaneously. However, given few new users, these WSPs may not offer any new plan, as the benefit from attracting new users cannot compensate for the immediate revenue loss. Perhaps surprisingly, we show that WSPs may not benefit from the option of innovative data plans, and their profits can decrease with new user number.
  \item \emph{Comparison between rollover and shared data plans:} In rollover data plan, a WSP's profit change is not affected by heavy users, while the timing of the shared data plan depends on the composition of light and heavy users in a family. In a saturated market with few new users, WSPs with similar market shares will not offer rollover data plans due to the great loss of overage charges, whereas they will still consider shared data plans based on the proportions of (heavy, heavy) and (heavy, light) families.
\end{itemize}

The rest of this paper is organized as follows. The system model and problem formulation is given in Section \ref{sec_model}.
In Section \ref{sec_rollover}, WSPs' optimal decisions under rollover data plan is analyzed and the equilibrium rollover time is presented. In Section \ref{sec_share}, we study the WSPs' equilibrium timing under shared data plan. Section \ref{sec_conclude} concludes this paper.

\section{System Model and Problem Formulation}\label{sec_model}


We consider a typical wireless data service market including competitive WSPs. For ease of exposition, we consider the two WSPs case as in \cite{musacchio2006game},\cite{duan2015economic}. Later in Section \ref{sec_rollover}, we also discuss the case with arbitrary number of symmetric WSPs. In reality, the data prices are similar for the WSPs with similar service quality. For example, in the USA, the subscription fee for $2$ GB data is $20$ USD per month for both Verizon and T-mobile. Thus, we assume that the WSPs offer the two-part tariff data plan $(P,B,p)$ with the same data price to their users, where a user is given a monthly data cap $B$ at a fixed lump-sum fee $P$, and should pay for extra data beyond $B$ at a costly unit price $p$. Without considering any future new user, there are $2N$ users currently and WSP $i, i\in\{1,2\}$ covers $2N_i$ users with current market share $\eta_i=N_i/N$.

In the following, we will first introduce the user model in the traditional data plan before the introduction of any innovative data plan. Then we will characterize the WSPs' pros and cons for upgrading traditional plans to rollover and shared data plans, and formulate the game theoretic models for the WSPs' timing competition. The key parameters which will be used throughout the paper are summarized in Table \ref{table1}.


\begin{table}[htbp]
\centering
\caption{Key Notations and Their Physical Meaning}\label{table1}
\begin{tabular}{|l|l|}
\hline \textbf{Symbol} & \textbf{Description}
\\\hline $B$ & subscribed data quota
\\\hline $P$ & subscription fee
\\\hline $p$ & unit price for extra data
\\\hline $\alpha$ & percentage of heavy users
\\\hline $D_l, D_h$ & maximum monthly data usage of light users,\\ & heavy users
\\\hline $d_l, d_h$ & minimum monthly data usage of light users,\\ & heavy users
\\\hline $u_l, u_h$ & monthly data usage of light users, heavy users
\\\hline $\eta_i$ & initial market share for WSP $i$
\\\hline $\eta_0$ & initial proportion of new users to existing users
\\\hline $\lambda_0, \lambda$ & new user arrival rate and churn rate
\\\hline $S$ & discount rate over time
\\\hline
\end{tabular}
\end{table}


In both the existing and new user pools, we classify the data users into two types according to their monthly data usage behaviors: $\alpha\in(0,1)$ percentage of heavy users who may exceed the data quota and $1-\alpha$ percentage of light users who will not exceed the data quota. For example, according to \cite{percentageHU}, more than $25\%$ of AT\&T customers (heavy users) paid an overage charge within six months. A light user's monthly data usage $u_l$ is a random variable on his possible usage range $d_l\leq u_l\leq D_l$, where $d_l$ and $D_l$ are the minimum and maximum monthly data usage, respectively. As a light users seldom exceeds the data cap $B$ (i.e., $D_l\leq B$), his expected monthly cost is $\bE C_l=P$. Similarly, a heavy user's monthly data usage $u_h$ is a random variable with probability density function $f_h(u_h)$ on his possible usage range $d_h\leq u_h\leq D_h$. In the existing two-part tariff, given $D_h>B$, a heavy user may overuse and his monthly cost by consuming $u_h$ is
\bee\label{equ_Ch} \setlength{\abovedisplayskip}{3pt}
\setlength{\belowdisplayskip}{3pt}
C_h=P+p(u_h-B)^+, \ene
where $(x)^+=\max(x,0)$.

By taking the expectation of (\ref{equ_Ch}) with respect to $u_h$, the heavy user's expected cost is
\bee \bE C_h=P+p\int_B^{D_h}(u_h-B)f_h(u_h)du_h. \ene

Take the uniform distribution as in \cite{duan2013economics} as an example, e.g., $f_h(u_h)=\frac{1}{D_h-d_h}$ for a heavy user, the heavy user's expected cost is
\bee\label{equ_ECh}\setlength{\abovedisplayskip}{3pt}
\setlength{\belowdisplayskip}{3pt}  \bE C_h=P+p\frac{(D_h-B)^2}{2D_h}. \ene

Note that the data usage distribution will only affect the users' expected cost reduction under innovative data plan and will not affect the WSPs' equilibrium timing analysis as shown in Sections \ref{sec_rollover} and \ref{sec_share}.

\newcounter{firstfig1}
\begin{figure*}[ht]
\setcounter{firstfig1}{\value{figure}}
\setcounter{figure}{1}
\centering
\subfigure[Phase I: $0\leq t\leq T_i$]{\label{PhaseI}
\begin{minipage}{.3\textwidth}
\includegraphics[width=1\textwidth]{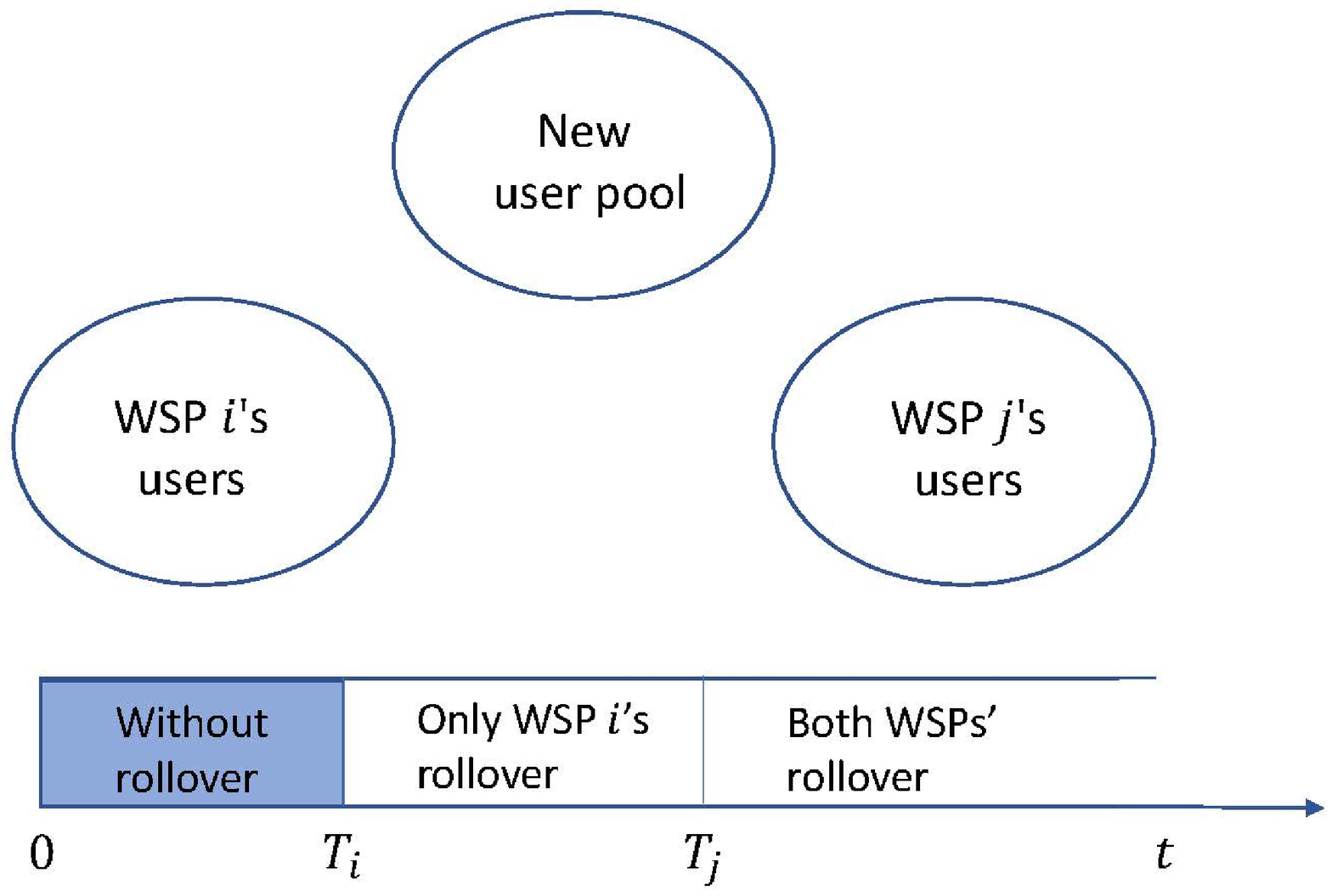}
\end{minipage}
}
\subfigure[Phase II: $T_i<t\leq T_j$]{\label{PhaseII}
\begin{minipage}{.3\textwidth}
\includegraphics[width=1\textwidth]{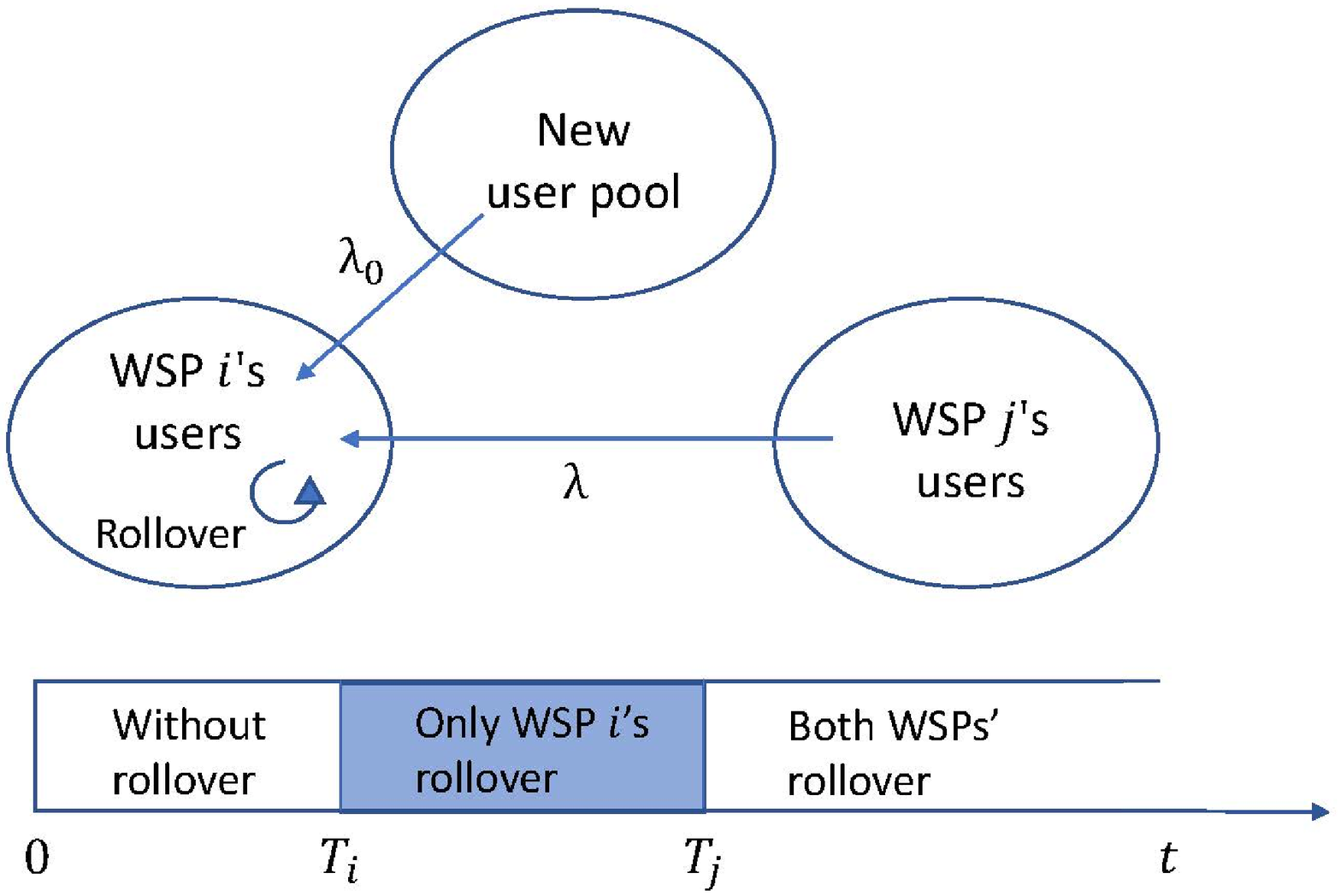}
\end{minipage}
}
\subfigure[Phase III: $t>T_j$]{\label{PhaseIII}
\begin{minipage}{.3\textwidth}
\includegraphics[width=1\textwidth]{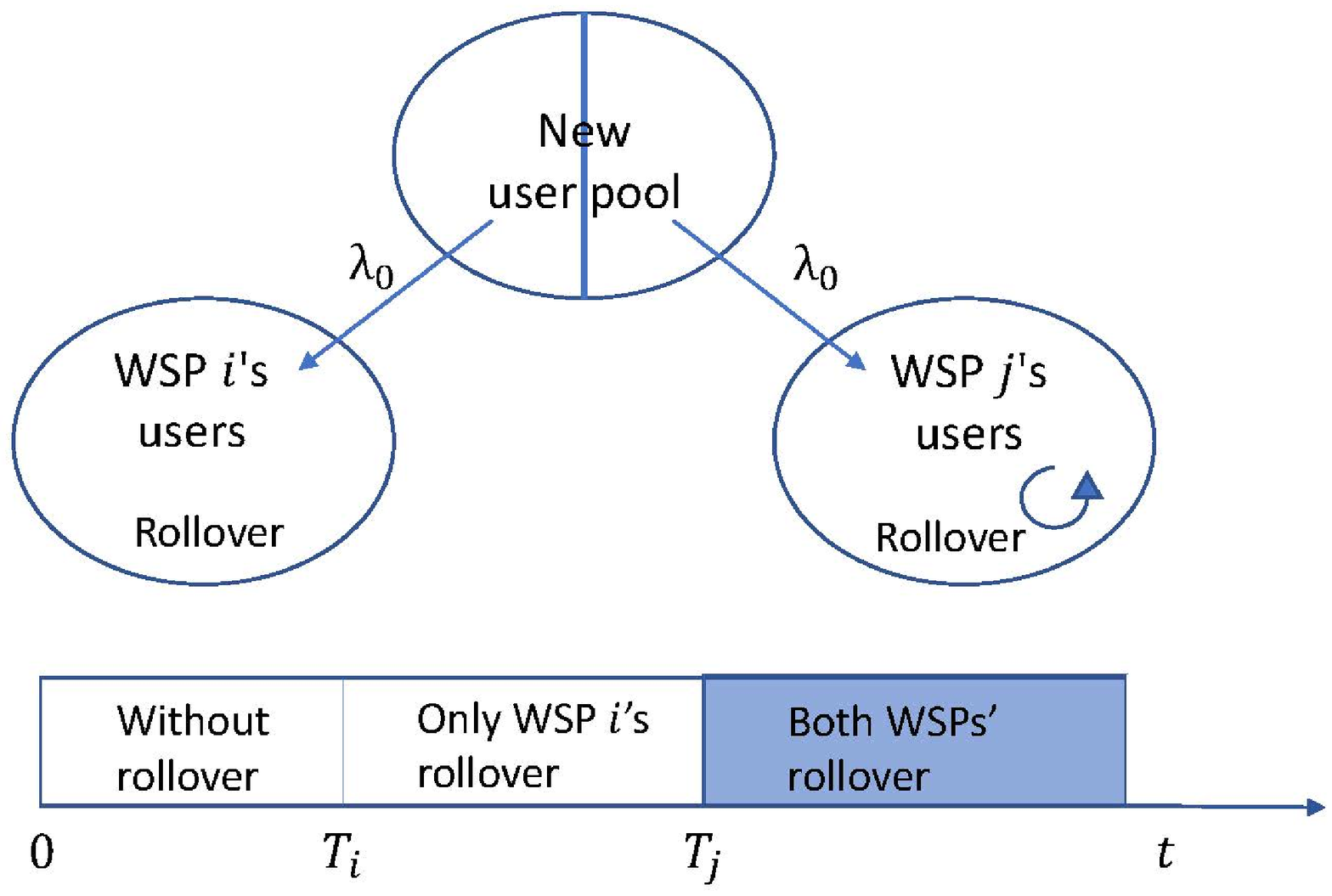}
\end{minipage}
}
\caption{Illustration of dynamic (heavy) user subscription when WSP $i$ chooses rollover upgrade earlier than WSP $j$ ($T_i<T_j$): Phase I without rollover data, Phase II with WSP $i$'s rollover service, Phase III with both WSPs' rollover services.}\label{userchurnrollover}
\end{figure*}

Besides the existing $2N$ users, there are potentially $2N_0$ new users and they can add proportion $\eta_0=N_0/N$ to the market, where $\eta_0$ may be even larger than 1. The number of new users is different across countries, e.g., large in China but small in USA. These new users will gradually join the market in the long run. Many of them are heavy users who are sensitive to the overage charge of the existing plan, but they will consider joining the innovative data plan (if any) for cost-saving. Besides the new user arrival, the existing users of a WSP will also consider switching to the other WSP's innovative data plan. Due to the switching cost $c_s$ for changing the subscribed WSP (e.g., contract termination fee),
the existing users will switch from one WSP to the other (first providing innovative data plan) gradually rather than all at once \cite{duan2015economic}.
As new user arrival or existing user churn between WSPs takes time,
we denote the new user arrival rate as $\lambda_0$ and existing user churn rate as $\lambda$, respectively, both of which depend on the user's cost reduction/saving $\Delta$ under the new innovative data plan. Note that we will estimate $\Delta$ according to the corresponding innovative data plan as specified later. If the cost reduction is large, the users will
switch to the innovative data plan with large rates $\lambda, \lambda_0$. Moreover, for the users who already subscribed to the WSP's
innovative data plan to enjoy the cost reduction $\Delta$, they will not switch to the other WSP even when the latter party also offers the
innovative data plan later due to the switching cost and lock-in effect.
This lock-in effect actually motivates the WSPs to compete to first offer innovative data plans.

Mathematically, we use differential equations to model the impact of switching cost $c_s$ and its resultant lock-in effect. Suppose WSP $i$ upgrades earlier than WSP $j$, i.e., $T_i<T_j$, WSP $j$'s left user share/proportion $\eta_j(t)$ is dynamically shrinking and follows
\bee\label{equ_dynamic_eta_i} \frac{d\eta_j(t)}{dt}=-\lambda(c_s,\Delta)\eta_j(t), t\in[T_i,T_j], \ene
where $\eta_j(t)$ starts with $\eta_j$ initially and $\lambda(c_s, \Delta)$ is the user churn rate. In general, $\lambda(c_s, \Delta)$
increases with cost reduction $\Delta$ under the new data plan, and decreases with $c_s$ which keeps users from churning. Note that the user churn rate is also affected by the difference between the two WSPs' wireless service qualities (e.g., signal coverage and the delay of data delivery), and the users will churn to the WSP with better service quality at a larger churn rate. Our analysis methods in Sections \ref{sec_rollover} and \ref{sec_share} still apply for the asymmetric churn rates case, though the analysis is more complicated with more cases. After $T_j$,
WSP $j$'s users also enjoy the cost reduction and will no longer churn to WSP $i$ to avoid switching cost. By solving (\ref{equ_dynamic_eta_i}), we have \bee \eta_j(t)=e^{-\lambda(c_s,\Delta)(t-T_i)}\eta_j, t\in[T_i,T_j]. \ene
Similar to \cite{duan2015economic}, we can validate the value of $\lambda$ by using real data.


Similarly, given the initial proportion of new users $\eta_0$, the dynamic of the left user proportion $\eta_0(t)$ of new users is
\bee\label{equ_dynamic_eta_0} \frac{d\eta_0(t)}{dt}=-\lambda_0(\Delta)\eta_0(t), t\in[T_i,\infty), \ene
where $\lambda_0(\Delta)$ increases with the cost reduction $\Delta$ and does not depend on $c_s$, since new users are not locked in any WSP initially. By solving (\ref{equ_dynamic_eta_0}), we have \bee \eta_0(t)=e^{-\lambda_0(\Delta)(t-T_i)}\eta_0, t\in[T_i,\infty). \ene


Then, for WSP $i$ who upgrades first at time $T_i$, its user share $\eta_i(t)$ at time $t$ increases from its initial market share $\eta_i$ and follows
\begin{align} \eta_i(t)=\eta_i+(1-e^{-\lambda(c_s,\Delta)(t-T_i)})\eta_j+(1-&e^{-\lambda_0(\Delta)(t-T_i)})\eta_0,\notag\\
&t\in[T_i,T_j] \end{align}


In the following, we will formulate the WSPs' timing problems for providing rollover and shared data plans, respectively.

\subsection{Problem Formulation for Rollover Data Plan}

To attract users from the new user pool and the other WSP, a WSP can provide rollover data plan at a proper time. Note that light users under-use the data quota and upgrading to rollover plan or not does not change their costs and subscription. The rollover upgrade from the existing plan only affects the heavy users' subscription (from the new user pool and the other WSP). Thus, we only need to consider the heavy users who can reduce their overage charges by using the rollover. We consider the typical rollover data plan (e.g., provided by AT\&T), where the WSP allows its users to keep unused data quota $(B-u_h(t-1))^+$ only from last month $t-1$ to next month $t$, and the rollover data is consumed only after the current month's data quota is used up. At time $t$, the heavy user's data quota now increases from $B$ to $B+(B-u_h(t-1))^+$ and his monthly cost by consuming random $u_h(t)$ in this month is
\bee\label{equ_Chrollover}\setlength{\abovedisplayskip}{3pt}
\setlength{\belowdisplayskip}{3pt}  C_h^r(t)=P+p\Big(u_h(t)-B-(B-u_h(t-1))^+\Big)^+. \ene

By taking expectation of (\ref{equ_Chrollover}) with respect to both $u_h(t)$ and $u_h(t-1)$, the heavy user's expected cost under the rollover data plan is
\bee\begin{split}\setlength{\abovedisplayskip}{3pt}
\setlength{\belowdisplayskip}{3pt}  \bE C_h^r=&P+p\int_{\xi}^{D_h}\Big(u_h(t)-\xi\Big)f_h(u_h(t))du_h(t). \end{split}\ene
where $\xi=B+\int_{d_h}^B(B-u_h(t-1))f_h(u_h(t-1))du_h(t-1)$.

Take the uniform distribution as an example, we derive the heavy user's expected cost under the rollover data plan as follows. 

\begin{lem} For the uniform distribution, the heavy user's expected cost under the rollover data plan is
\begin{equation}\label{equ_EChr}
\setlength{\abovedisplayskip}{3pt}
\setlength{\belowdisplayskip}{3pt}
\bE C_h^r=\left\{
\begin{array}{l}
P+p\frac{2(D_h-B)^3}{3D_h^2}, \text{if $B<D_h\leq 2B$;} \\
P+p(\frac{2(D_h-B)^3}{3D_h^2}-\frac{(D_h-2B)^3}{6D_h^2}),  \text{if $D_h>2B$,}\\
\end{array}
\right.
\end{equation}
which is smaller than traditional cost $\bE C_h$ in (\ref{equ_ECh}).
\end{lem}

\newcounter{my10}
\begin{figure*}[ht]
\setcounter{my10}{\value{equation}}
\setcounter{equation}{12}
\footnotesize{\bee\begin{split}\label{equ_R_i^r_early} R_i^{r,\leq}=&\int_0^{T_i}2\alpha N_i\bE C_he^{-St}dt
+\int_{T_i}^{T_j} \Big(2\alpha N_j(1-e^{-\lambda (t-T_i)})\bE C_h^r+2\alpha N_i\bE C_h^r
+2\alpha N_0(1-e^{-\lambda_0 (t-T_i)})\bE C_h^r\Big) e^{-St}dt\\
+&\int_{T_j}^\infty\Big(2\alpha N_j(1-e^{-\lambda (T_j-T_i)})\bE C_h^r+2\alpha N_i\bE C_h^r
+\alpha N_0e^{-\lambda_0 (T_j-T_i)}(1-e^{-\lambda_0 (t-T_j)})\bE C_h^r
+2\alpha N_0(1-e^{-\lambda_0 (T_j-T_i)})\bE C_h^r\Big)e^{-St}dt. \end{split}\ene
\bee\begin{split}\label{equ_R_i^r_late}\setlength{\abovedisplayskip}{3pt}
\setlength{\belowdisplayskip}{3pt}
R_i^{r,>}=&\int_0^{T_j}2\alpha N_i\bE C_he^{-St}dt
+\int_{T_j}^{T_i} 2\alpha N_ie^{-\lambda (t-T_j)}\bE C_h e^{-St}dt
+\int_{T_i}^\infty \Big(2\alpha N_ie^{-\lambda(T_i-T_j)}
+\alpha N_0e^{-\lambda_0 (T_i-T_j)}(1-e^{-\lambda_0 (t-T_i)})\Big)\bE C_h^re^{-St}dt. \end{split}\ene
}
\setcounter{equation}{\value{my10}}
\hrulefill
\end{figure*}

\newcounter{firstfig}
\begin{figure}[ht]
\setcounter{firstfig}{\value{figure}}
\setcounter{figure}{0}
\centering\includegraphics[scale=0.5]{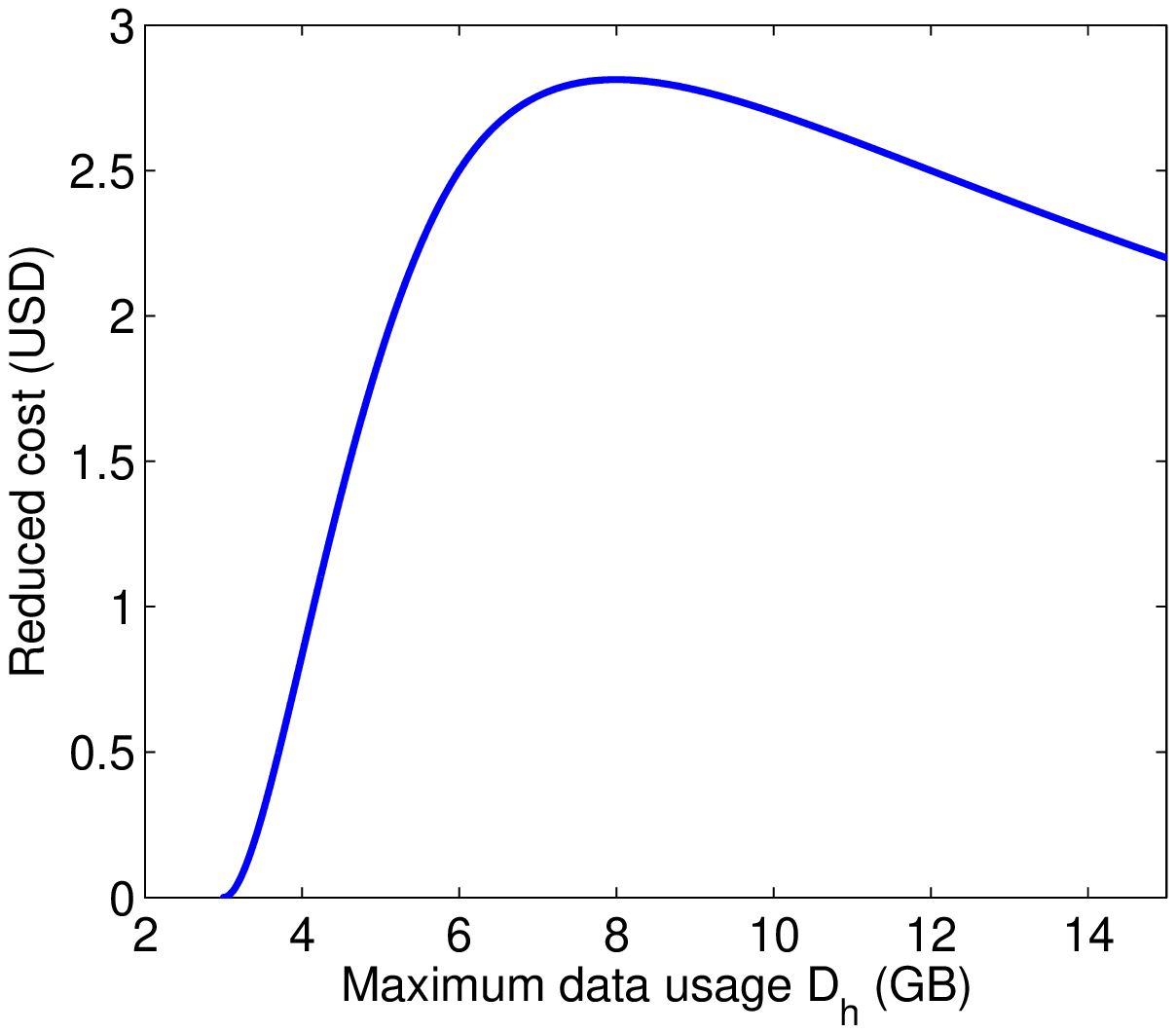}\caption{Heavy user's expected cost reduction $\bE C_h-\bE C_h^r$ vs his maximum data usage $D_h$ when $B=3$ GB, $p=10$ USD/GB.}\label{reduceC}
\end{figure}

\newcounter{firstfig3}
\setcounter{firstfig3}{\value{figure}}
\setcounter{figure}{2}

As shown in Fig. \ref{reduceC}, the cost reduction $\bE C_h-\bE C_h^r$ after adding rollover data first increases with the maximum data usage $D_h$ as the heavy user is more likely to rollover data and save cost. The cost reduction then decreases as there is less left data to rollover.


Due to the cost saving, heavy users will gradually churn to the WSP with rollover data upgrade. Fig. \ref{userchurnrollover} illustrates how existing heavy users switch from WSP $i$ to WSP $j$ and how new (heavy) users choose WSPs over the time horizon $t$. Time $0$ means the earliest possible time for data plan upgrade. Here we suppose WSP $i$ upgrades earlier than WSP $j$ (i.e., $T_i\leq T_j$) in Fig. \ref{userchurnrollover}.\footnote{If $T_i=T_j$, we do not have Phase II in Fig. \ref{userchurnrollover}.} As explained earlier, we do not consider existing and new light users, as they are not affected by the rollover data plan and do not matter the WSPs' upgrade timing. There are three phases for dynamic user subscription:
\begin{itemize}
  \item Phase I ($0\leq t\leq T_i$) as in Fig. \ref{PhaseI}. No WSP offers the rollover data plan. Thus no existing heavy user switches between WSPs and no new heavy user joins any network.
  \item Phase II ($T_i<t\leq T_j$) as in Fig. \ref{PhaseII}. WSP $i$ offers the rollover data plan at time $T_i$ but WSP $j$ has not. The existing heavy users of WSP $i$ benefit from the upgrade to rollover data plan immediately from time $T_i$, and the heavy users of WSP $j$ and the new user pool gradually join WSP $i$ at rates $\lambda$ and $\lambda_0$, respectively. The number of heavy users switched from WSP $j$ to $i$ from time $T_i$ to $t$ is $2\alpha N_j(1-e^{-\lambda (t-T_i)})$, where there are originally $2\alpha N_j$ users in WSP $j$. The number of new heavy users joining WSP $i$ is $2\alpha N_0(1-e^{-\lambda_0 (t-T_i)})$.
  \item Phase III ($t>T_j$) as in Fig. \ref{PhaseIII}. Both WSPs offer rollover data plans, and no existing users will switch to the other WSP. The new heavy users are equally likely to subscribe to both WSPs (for cost saving) since $T_j$. There are only $2\alpha N_0e^{-\lambda_0(T_j-T_i)}$ left in the new heavy user pool at time $T_j$, and the numbers of new heavy users subscribed to each WSP from time $T_j$ to $t$ are the same, i.e., $\alpha N_0 e^{-\lambda_0 (T_j-T_i)}(1-e^{-\lambda_0 (t-T_j)})$.
\end{itemize}

Based on the dynamic subscription of users over the three phases, we are ready to characterize the profits of WSPs. In reality, each WSP values its future profit less importantly due to depreciation. We denote the discount rate over time as $S>0$, and model the discount factor till time $t$ as $e^{-St}$ as in \cite{chiang1984fundamental}. Without offering rollover data plan (i.e., $T_i=T_j=\infty$), the long-term profit of WSP $i, i=1,2$, from its existing heavy users is
\bee\label{equ_Ribefore}\setlength{\abovedisplayskip}{3pt}
\setlength{\belowdisplayskip}{3pt}  R_i=\int_0^{\infty}2\alpha N_i\bE C_h e^{-St}dt=2\alpha N_i\bE C_h\frac{1}{S}, \ene
where $\bE C_h$ is given in (\ref{equ_ECh}).

\begin{figure*}[t]
\centering
\subfigure[Phase I: $0\leq t\leq T_i$]{\label{PhaseIshare}
\begin{minipage}{.3\textwidth}
\includegraphics[width=1\textwidth]{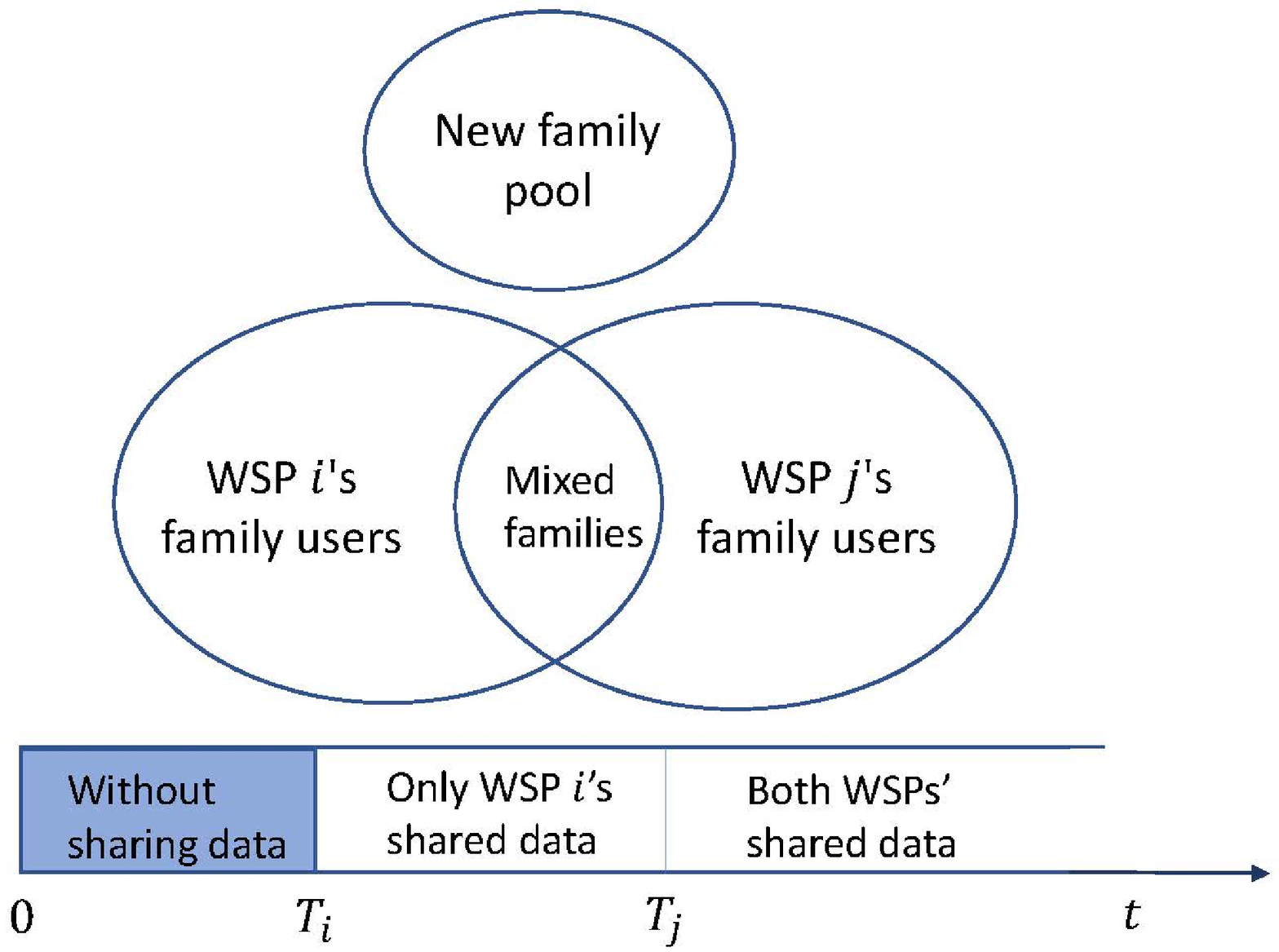}
\end{minipage}
}
\subfigure[Phase II: $T_i<t\leq T_j$]{\label{PhaseIIshare}
\begin{minipage}{.3\textwidth}
\includegraphics[width=1\textwidth]{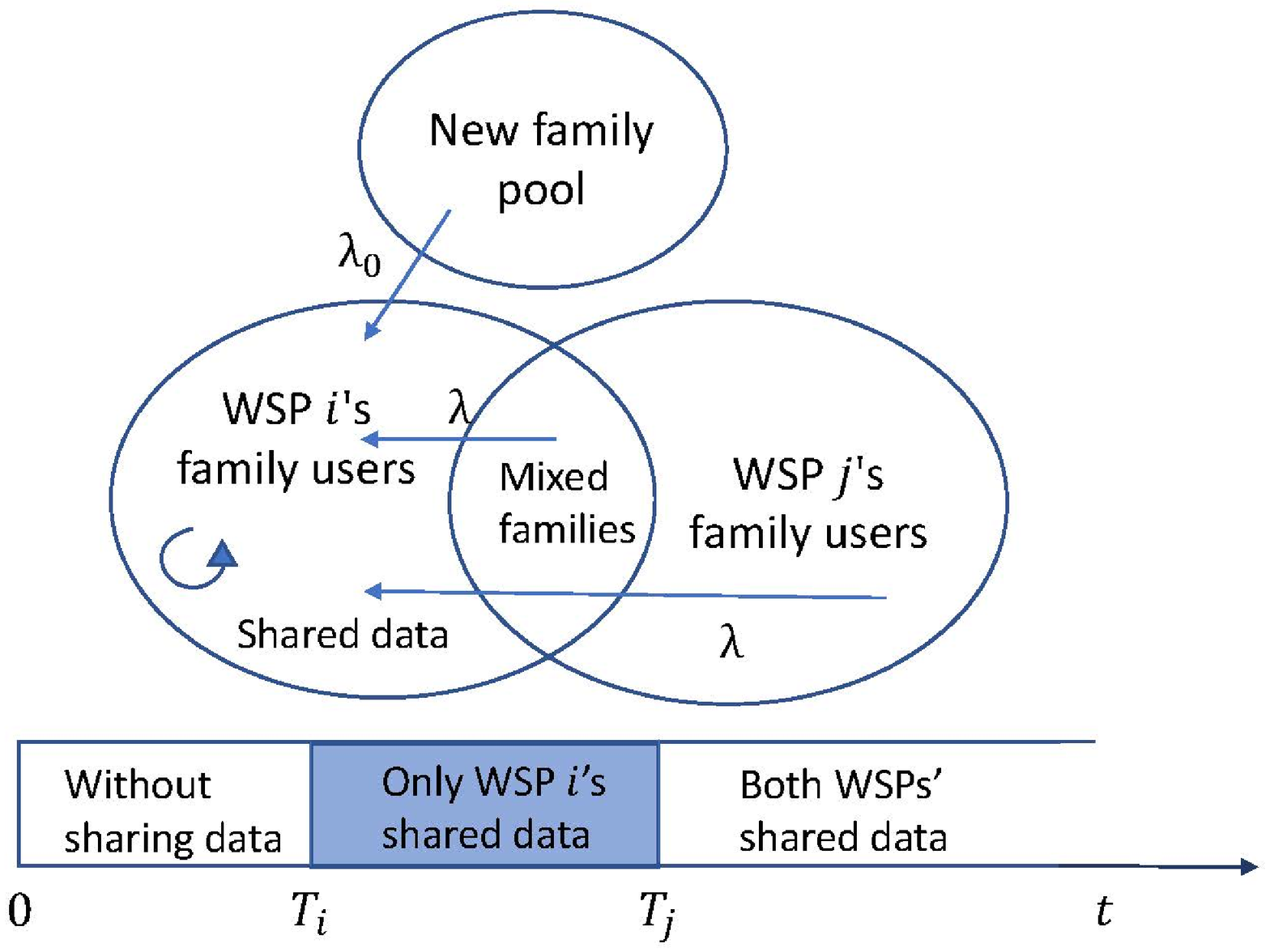}
\end{minipage}
}
\subfigure[Phase III: $t>T_j$]{\label{PhaseIIIshare}
\begin{minipage}{.3\textwidth}
\includegraphics[width=1\textwidth]{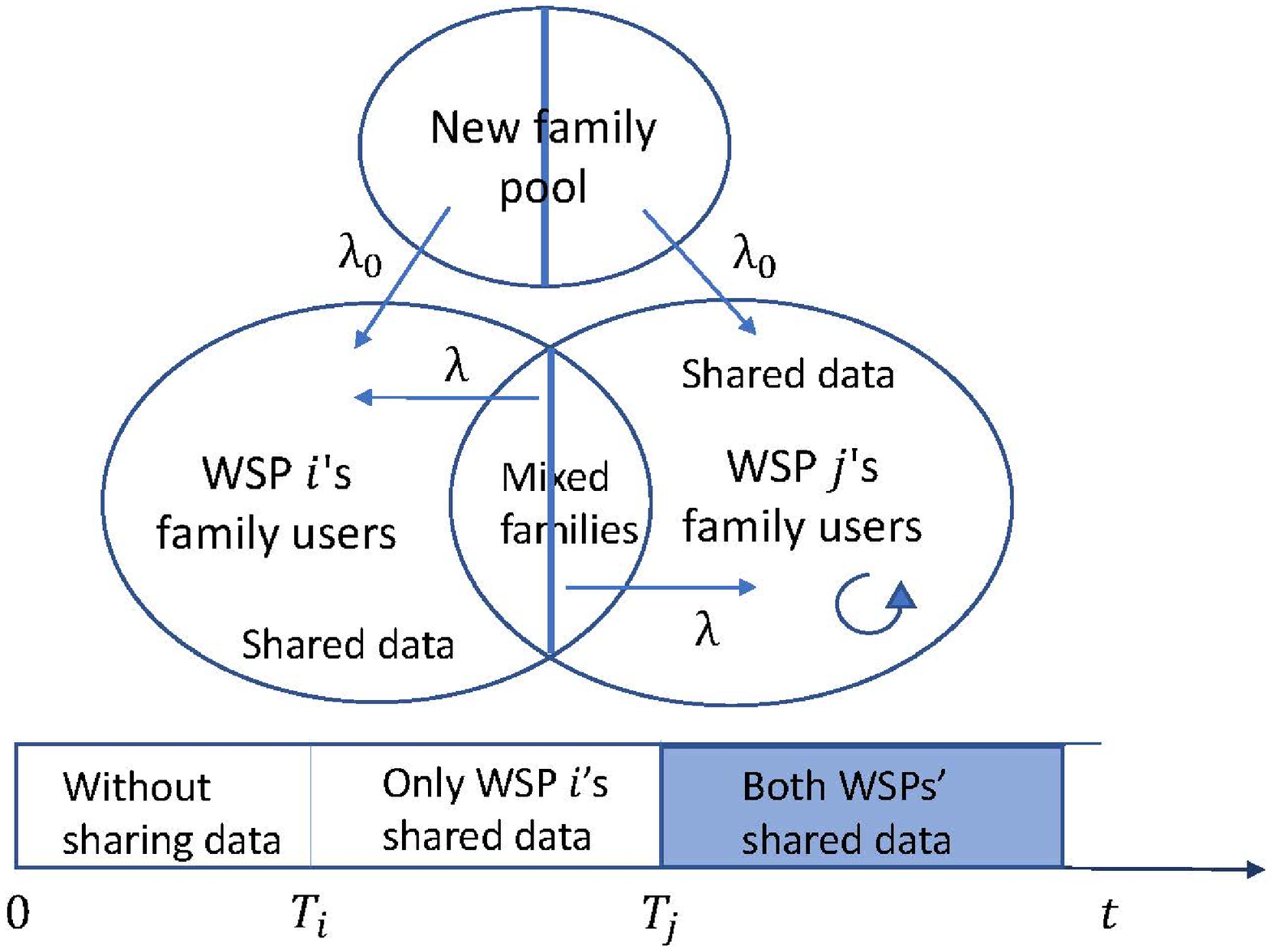}
\end{minipage}
}
\caption{Illustration of (heavy, heavy) and (heavy, light) families subscription under WSPs' shared data plans when WSP $i$ upgrades earlier than WSP $j$ ($T_i<T_j$): Phase I without shared data, Phase II with WSP $i$'s shared data service, Phase III with both WSPs' shared data services.}\label{userchurnshare}
\end{figure*}

Given WSP $j$'s rollover time $T_j\geq 0$, WSP $i$ can choose to rollover earlier or later than WSP $j$. As shown in Fig. \ref{userchurnrollover}, WSP $i$'s long-term profits through the three time phases for $T_i\leq T_j$ and $T_i>T_j$ are given in (\ref{equ_R_i^r_early}) and (\ref{equ_R_i^r_late}), respectively. Based on these results, we describe the non-cooperative timing game for rollover data plan as follows.
\begin{itemize}
  \item Players: WSPs $1$ and $2$.
  \item Strategy spaces: WSP $i\in\{1,2\}$ can choose rollover time $T_i$ from the feasible set $\cT_i=[0,\infty]$.
  \item Payoff functions: WSP $i\in\{1,2\}$ wants to maximize its profit defined in (\ref{equ_R_i^r_early}) and (\ref{equ_R_i^r_late}).
\end{itemize}
\newcounter{my11}
\setcounter{my11}{\value{equation}}
\setcounter{equation}{14}

\newcounter{my12}
\begin{figure*}[ht]
\setcounter{my12}{\value{equation}}
\setcounter{equation}{21}
\footnotesize{\bee\label{equ_R_se}\setlength{\abovedisplayskip}{3pt}
\setlength{\belowdisplayskip}{3pt} \begin{split} R_i^{s,\leq}=&\Big(\cE N(\frac{1}{S}-\frac{1}{\lambda+S})+\eta_i^2N\frac{1}{\lambda+S}\cE-\eta_i N\cD\frac{1}{S}
+\eta_i\eta_jN\cD\frac{1}{\lambda+S}+N_0\cE(\frac{1}{S}-\frac{1}{\lambda_0+S})\Big)e^{-ST_i}
-(1-\eta_i)N\cE(\frac{1}{S}-\frac{1}{\lambda+S})e^{-(\lambda+S)T_j+\lambda T_i}\\
&-\frac{1}{2}N_0\cE(\frac{1}{S}-\frac{1}{\lambda_0+S})e^{-(\lambda_0+S)T_j+\lambda_0 T_i}+\eta_i N\cD\frac{1}{S}, \end{split}\ene
\bee\label{equ_R_sl}\setlength{\abovedisplayskip}{3pt}
\setlength{\belowdisplayskip}{3pt} \begin{split} R_i^{s,>}=&\eta_iN\cD\frac{1}{S}-\eta_iN\cD(\frac{1}{S}-\frac{1}{\lambda+S})e^{-ST_j}
+\eta_i N(\cE\frac{1}{S}-\eta_j \cE\frac{1}{\lambda+S}
-\eta_i \cD\frac{1}{\lambda+S})e^{-(\lambda+S)T_i+\lambda T_j}
+\frac{1}{2}N_0\cE(\frac{1}{S}-\frac{1}{\lambda_0+S})e^{-(\lambda_0+S)T_i+\lambda_0 T_j}. \end{split}\ene}
\setcounter{equation}{\value{my12}}
\hrulefill
\end{figure*}



\subsection{Problem Formulation for Shared Data Plan}

To attract more subscribers, a WSP can also choose to provide shared data plan at a proper time. We assume the average family size for choosing a shared data plan is two, and our analysis can also be extended to the case of three or more members in an average family. By combining two individual users' data cap $B$ and lump sum fee $P$, the shared data plan is $(2P, 2B, p)$ for the whole family. Suppose two users (no matter heavy or light users) form a family randomly, and each user is a heavy user with probability $\alpha$. Based on different combinations of heavy and light users, the proportions of (heavy, heavy), (heavy, light) and (light, light) families are $\alpha^2, 2\alpha(1-\alpha)$ and $(1-\alpha)^2$, respectively. Compared to the rollover data plan where a WSP's profit change is only affected by heavy users, the timing of the shared data plan here depends on the composition of light and heavy users in a family. A light user in a (heavy, light) family can also change to the WSP with shared data plan to reduce the cost of the heavy user in the same family. We note that in each of the (light, light) families, the two users do not benefit from the shared data plan. This type of families does not affect the WSPs' profit change after the shared data plan, and we only need to consider the (heavy, heavy) and (heavy, light) families when deciding the timing of offering shared data plan. Similarly, we do not consider bachelordom users without family. For each type of families, the proportion of both family users choosing the same WSP $i$ of market share $\eta_i$ is $\eta_i^2$, and we call such families as pure families. Similarly, the proportion of each user choosing a different WSP is $2\eta_i \eta_j$, and we call such families as mixed families.


Under the shared data plan, the family's monthly cost is
\bee\label{equ_Cs}\setlength{\abovedisplayskip}{3pt}
\setlength{\belowdisplayskip}{3pt} C_s=2P+p(u-2B)^+,\ene
where random variable $u$ is the total family data usage: $u=u_h+u_l$ for a (heavy, light) family and $u=u_h+u_h$ for a (heavy, heavy) family.

By taking the expectation of (\ref{equ_Cs}) with respect to $u$, the expected costs of (heavy, light) and (heavy, heavy) families are
\bee \setlength{\abovedisplayskip}{3pt}
\setlength{\belowdisplayskip}{3pt} \bE C_{h,l}=2P+p\int_{2B}^{D_l+D_h}(u-2B)f(u)du, \ene
and
\bee \setlength{\abovedisplayskip}{3pt}
\setlength{\belowdisplayskip}{3pt} \bE C_{h,h}=2P+p\int_{2B}^{D_h+D_h}(u-2B)f(u)du. \ene

Take the uniform distribution of users' data usage as an example, the distribution of $u$ for a (heavy, light) family is
\begin{equation}\label{equ_disu}
\setlength{\abovedisplayskip}{3pt}
\setlength{\belowdisplayskip}{3pt}
f(u)=\left\{
\begin{array}{l}
\frac{u}{D_l D_h}, \text{~~~~~~~~~~~~~~if $0\leq u\leq D_l$;} \\
\frac{1}{D_h}, \text{~~~~~~~~~~~~~~~~~if $D_l< u\leq D_h$;} \\
\frac{D_l+D_h-u}{D_l D_h}, \text{~~~~~~~~~if $D_h<u\leq D_l+D_h$.} \\
\end{array}
\right.
\end{equation}

For a (heavy, heavy) family, its distribution of $u$ is the similar as (\ref{equ_disu}) by letting $D_l=D_h$. Then, we obtain the expected costs of (heavy, heavy) and (heavy, light) families as follows.

\begin{lem} For the uniform distribution of users' data usage, the (heavy, heavy) family's expected cost is
\begin{equation}
\setlength{\abovedisplayskip}{3pt}
\setlength{\belowdisplayskip}{3pt}
\bE C_{h,h}=\left\{
\begin{array}{l}
2P+\frac{4p}{3D_h^2}(D_h-B)^3, \text{~~~~~~~~~~~~~if $D_h\leq 2B$;} \\
2P+\frac{p}{D_h^2}(D_h^3-2BD_h^2+\frac{4}{3}B^3), \text{~if $D_h>2B$.} \\
\end{array}
\right.
\end{equation}
and the (heavy, light) family's expected cost is
\begin{equation}\begin{split}
\setlength{\abovedisplayskip}{3pt}
\setlength{\belowdisplayskip}{3pt}
\bE C_{h,l}=\left\{
\begin{array}{l}
2P+\frac{p}{6D_h D_l}(D_h+D_l-2B)^3, \text{~if $D_h\leq 2B$;} \\
2P+\frac{p}{D_h}(\frac{D_h^2}{2}-2BD_h+2B^2\\
+\frac{D_l^2}{6}+\frac{D_h D_l}{2}-BD_l), \text{~~~~~~~~~~if $D_h>2B$.} \\
\end{array}
\right.
\end{split}\end{equation}
\end{lem}


Due to the overage cost saving, (heavy, heavy) and (heavy, light) families will gradually churn to the WSP with shared data upgrade. Fig. \ref{userchurnshare} illustrates how existing (heavy, heavy) and (heavy, light) families switch from one WSP to another and how new (heavy, heavy) and (heavy, light) families choose WSPs in the time horizon $t$. Here we suppose WSP $i$ upgrades earlier than WSP $j$ (i.e., $T_i\leq T_j$) in Fig. \ref{userchurnshare} and can simply skip Phase II if $T_i=T_j$. As explained earlier, here we do not count existing and new (light, light) families, as they are not affected by the shared data plan and do not matter the timing decisions.
\begin{itemize}
  \item Phase I ($0\leq t\leq T_i$) as in Fig. \ref{PhaseIshare}. No WSP offers the shared data plan. Thus, no existing user switches between WSPs and no new user joins any network.
  \item Phase II ($T_i<t\leq T_j$) as in Fig. \ref{PhaseIIshare}. WSP $i$ offers the shared data plan at time $T_i$ but WSP $j$ has not. If both users in a family have subscribed to WSP $i$, they can upgrade to the shared data plan immediately at time $T_i$. Otherwise, the existing families (with at least one family member in WSP $j$) and the new families will join WSP $i$ gradually at churn rates $\lambda$ and $\lambda_0$, respectively. Thus, the subscribed families to WSP $i$ from $T_i$ to $t$ are: (i) $\eta_i^2\alpha^2 N$ pure (heavy, heavy) families from WSP $i$; (ii) $\eta_j^2 \alpha^2 N(1-e^{-\lambda (t-T_i)})$ pure (heavy, heavy) families from WSP $j$, where there are originally $\eta_j^2 \alpha^2 N$ pure (heavy, heavy) families in WSP $j$; (iii) $2\eta_i\eta_j\alpha^2 N(1-e^{-\lambda (t-T_i)})$ mixed (heavy, heavy) families; (iv) $2\eta_i^2\alpha(1-\alpha) N$ pure (heavy, light) families from WSP $i$; (v) $2\eta_j^2\alpha(1-\alpha)N(1-e^{-\lambda (t-T_i)})$ pure (heavy, light) families from WSP $j$; (vi) $4\eta_i\eta_j\alpha(1-\alpha)N(1-e^{-\lambda (t-T_i)})$ mixed (heavy, light) families; (vii) $\alpha^2 N_0(1-e^{-\lambda_0 (t-T_i)})$ new (heavy, heavy) families; (vii) $2\alpha(1-\alpha) N_0(1-e^{-\lambda_0 (t-T_i)})$ new (heavy, light) families.
  \item Phase III ($t>T_j$) as in Fig. \ref{PhaseIIIshare}. Both WSPs offer the shared data plan, and no pure families will switch to the other WSP. The mixed families and the new families are equally likely to subscribe to the two WSPs. Thus, a number $\eta_i\eta_j\alpha^2 Ne^{-\lambda (T_j-T_i)}(1-e^{-\lambda (t-T_j)})$ of mixed (heavy, heavy) families and a number $2\eta_i\eta_j\alpha(1-\alpha)Ne^{-\lambda (T_j-T_i)}(1-e^{-\lambda (t-T_j)})$ of mixed (heavy, light) families will subscribe to each WSP from time $T_j$ to $t$. And the number of new (heavy, heavy) families and (heavy, light) families subscribed to each WSP from time $T_j$ to $t$ are $\frac{1}{2}\alpha^2 N_0e^{-\lambda_0(T_j-T_i)}(1-e^{-\lambda_0 (t-T_j)})$ and $\alpha(1-\alpha) N_0e^{-\lambda_0(T_j-T_i)}(1-e^{-\lambda_0 (t-T_j)})$, respectively.
\end{itemize}

\newcounter{my1}
\begin{figure*}[ht]
\setcounter{my1}{\value{equation}}
\setcounter{equation}{26}
\begin{footnotesize}
\bee\label{equ_big1}\setlength{\abovedisplayskip}{3pt}
\setlength{\belowdisplayskip}{3pt}\begin{split} &\Big(2\alpha(N_i\bE C_h-N\bE C_h^r+N_j\frac{S}{\lambda+S}\bE C_h^r)
-2\alpha N_0\frac{\lambda_0}{\lambda_0+S}\bE C_h^r\Big)e^{-ST_i}
-2\alpha N_j(\frac{\lambda}{S}-\frac{\lambda}{\lambda+S})\bE C_h^re^{-ST_j}e^{-\lambda(T_j-T_i)}\\
&-\alpha N_0(\frac{\lambda_0}{S}-\frac{\lambda_0}{\lambda_0+S})\bE C_h^re^{-ST_j}e^{-\lambda_0(T_j-T_i)}=0. \end{split}\ene\end{footnotesize}
\setcounter{equation}{\value{my1}}
\hrulefill
\end{figure*}


Based on the dynamic subscription of users over the three phases, we can characterize the profits of WSPs as follows. Without offering shared data plan (i.e., $T_i=T_j=\infty$), the long-term profit of WSP $i, i=1,2$, from its existing (heavy, heavy) and (heavy, light) families is
\bee\begin{split}\setlength{\abovedisplayskip}{3pt}
\setlength{\belowdisplayskip}{3pt} R_i^s=&\int_{0}^{\infty}(2\alpha \eta_i N\bE C_h+2\eta_i N\alpha(1-\alpha)\bE C_l)e^{-St}dt\\
=&(2\alpha \eta_i N\bE C_h+2\eta_i N\alpha(1-\alpha)\bE C_l)\frac{1}{S}. \end{split}\ene

\newcounter{my13}
\setcounter{my13}{\value{equation}}
\setcounter{equation}{23}

WSP $i$'s long-term profits through the three time phases for $T_i\leq T_j$ and $T_i>T_j$ cases are given in (\ref{equ_R_se}) and (\ref{equ_R_sl}), respectively, where \bee\label{equ_cD}\setlength{\abovedisplayskip}{3pt}
\setlength{\belowdisplayskip}{3pt} \cD=2\alpha\bE C_h+2\alpha(1-\alpha)\bE C_l, \ene
\bee\label{equ_cE}\setlength{\abovedisplayskip}{3pt}
\setlength{\belowdisplayskip}{3pt} \cE=\alpha^2\bE C_{h,h}+2\alpha(1-\alpha)\bE C_{h,l}. \ene



Similar to the timing game of rollover data plan, we use a noncooperative game to model the interactions between the two WSPs, where each of them seeks to maximize its profit defined in (\ref{equ_R_se}) and (\ref{equ_R_sl}) by choosing the optimal timing of offering the shared data plan. In the following, we will analyze the timing equilibrium for the rollover data plan in Section \ref{sec_rollover} and shared data plan in Section \ref{sec_share}. 

\section{Timing Equilibrium for rollover data plan}\label{sec_rollover}

By announcing the rollover data plan, a WSP loses its immediate revenue from its existing users as the overage charge reduces (from $\bE C_h$ in (\ref{equ_ECh}) to $\bE C_h^r$ in (\ref{equ_EChr})), but it can gradually increase profit by attracting new users and those from the other WSP. Note that as existing users are mainly adults and potential new users could be teenagers who take time to mature, we expect $\lambda_0<\lambda$ in practice. In this section, we first analyze WSP $i$'s best rollover timing given any WSP $j$'s rollover time $T_j$, where $i, j\in\{1, 2\}$ and $i\neq j$. Then, we derive the equilibrium rollover timing of both WSPs.



By examining the first-order conditions of WSP's profit before and after its competitor's rollover, we derive the conditions that WSP $i$ will rollover first.

\begin{lem}\label{lem_rollfirst} Given WSP $j$'s rollover time $T_j$, WSP $i$ will upgrade to rollover data plan earlier than WSP $j$ ( i.e., $T_i\leq T_j$), if and only if one of the following conditions holds:
\begin{description}
         \item[(I)] Reduction of overage charge after rollover is mild, i.e., $\kappa_i\leq 1$, where  \bee\setlength{\abovedisplayskip}{3pt}
\setlength{\belowdisplayskip}{3pt} \kappa_i=\frac{2 \eta_i(\bE C_h-\bE C_h^r\frac{\lambda+S}{S})}{\eta_0\bE C_h^r\frac{\lambda_0}{S}}. \ene
         \item[(II)] $\kappa_i>1$ yet earlier upgrade's gain in market share still exceeds the later upgrade's benefit of overage charges (i.e., $R_i^{r,\leq}(\hat{T}_i)$ in (\ref{equ_R_i^r_early}) is larger than $R_i^{r,>}(T_j+\frac{\log\kappa_i}{\lambda-\lambda_0})$ in (\ref{equ_R_i^r_late})). Here, WSP $i$'s best upgrade time after $T_j$ is $T_j+\frac{\log\kappa_i}{\lambda-\lambda_0}$, and the best time before $T_j$ is $\hat{T}_i=\arg\max(R_i^{r,\leq}(\tilde{T}_i),R_i^{r,\leq}(0),R_i^{r,\leq}(T_j))$, where $\tilde{T}_i$ is the solution to (\ref{equ_big1}).
  \newcounter{my2}
\setcounter{my2}{\value{equation}}
\setcounter{equation}{27}
\end{description}
\end{lem}

\textbf{Proof:} If $\kappa_i\leq 1$, we have $\frac{\partial R_i^{r,>}(T_i,T_j)}{\partial T_i}\leq 0$. Thus, WSP $i$ will not offer rollover later than WSP $j$.

If $\kappa_i>1$, WSP $i$ will choose its upgrade time by comparing its profits before and after WSP $j$'s rollover. Note that $\frac{\partial^2 R_i^{r,>}(T_i,T_j)}{\partial T_i^2}<0$. By letting $\frac{\partial R_i^{r,>}(T_i,T_j)}{\partial T_i}=0$, we have WSP $i$'s best upgrade time after $T_j$ as $T_j+\frac{\log\kappa_i}{\lambda-\lambda_0}$. Then, we derive the optimal time before $T_j$. Denote $\tilde{T}_i$ as the solution to $\frac{\partial R_i^{r,\leq}(T_i,T_j)}{\partial T_i}=0$, which can be written as (\ref{equ_big1}). Note that $\tilde{T}_i$ should be less than $T_j$ and larger than $0$. Thus, by comparing the profits of extreme and boundary points, we have the best time before $T_j$ as $\hat{T}_i=\arg\max(R_i^{r,\leq}(\tilde{T}_i),R_i^{r,\leq}(0),R_i^{r,\leq}(T_j))$. Therefore, WSP $i$ will upgrade to rollover data plan earlier than WSP $j$ if $R_i^{r,\leq}(\hat{T}_i)\geq R_i^{r,>}(T_j+\frac{\log\kappa_i}{\lambda-\lambda_0})$. \qed

According to Lemma \ref{lem_rollfirst}, WSP $i$'s best response can be derived immediately in the following proposition.

\begin{pro}\label{pro_bestrollovertime} Given WSP $j$'s rollover time $T_j$, WSP $i$'s best rollover time $T_i^*(T_j)$ is
\begin{itemize}
\item If $T_j=0$, then
  \begin{equation}
  \setlength{\abovedisplayskip}{3pt}
\setlength{\belowdisplayskip}{3pt}
  T_i^*(T_j)=\left\{
  \begin{array}{l}
  0, \text{~~~~~if $\kappa_i\leq 1$;} \\
  \frac{\log\kappa_i}{\lambda-\lambda_0}, \text{if $\kappa_i>1$.} \\
  \end{array}
  \right.
  \end{equation}
\item If $T_j>0$, then
  \begin{equation}
  \setlength{\abovedisplayskip}{3pt}
\setlength{\belowdisplayskip}{3pt}
  T_i^*(T_j)=\left\{
  \begin{array}{l}
  \hat{T}_i, \text{~~~~~~~~~~~if (I) or (II) holds;} \\
  T_j+\frac{\log\kappa_i}{\lambda-\lambda_0}, \text{~Otherwise.} \\
  \end{array}
  \right.
  \end{equation}
\end{itemize}
\end{pro}

%

\newcounter{myver}
\begin{figure*}[ht]
\setcounter{myver}{\value{equation}}
\setcounter{equation}{29}
\footnotesize{\bee\label{equ_phi0}\begin{split} &\alpha N\bE C_h-2\alpha N\bE C_h^r+\alpha N\frac{S}{\lambda+S}\bE C_h^r
   -2\alpha \eta_0 N\frac{\lambda_0}{\lambda_0+S}\bE C_h^r
   -\alpha N(\frac{\lambda}{S}-\frac{\lambda}{\lambda+S})\bE C_h^r\Big(\frac{\eta_0\bE C_h^r\frac{\lambda_0}{S}}{\bE C_h-\bE C_h^r\frac{\lambda+S}{S}}\Big)^{\frac{\lambda+S}{\lambda-\lambda_0}}\\
   &+\alpha \eta_0N(\frac{\lambda_0}{\lambda_0+S}-\frac{\lambda_0}{S})\bE C_h^r\Big(\frac{\eta_0\bE C_h^r\frac{\lambda_0}{S}}{\bE C_h-\bE C_h^r\frac{\lambda+S}{S}}\Big)^{\frac{\lambda_0+S}{\lambda-\lambda_0}}=0,
   \end{split}\ene}
   \footnotesize{
\bee\label{equ_biglower} \underline{\eta_r}=\max\bigg(1-\tilde{\eta}_r, \min\Big(\max(1-\bar{\eta}_r, \frac{N_0\bE C_h^r\frac{\lambda_0}{S}}{2N(\bE C_h-\bE C_h^r\frac{\lambda+S}{S})}), \min(\bar{\eta}_r, 1-\frac{N_0\bE C_h^r\frac{\lambda_0}{S}}{2N(\bE C_h-\bE C_h^r\frac{\lambda+S}{S})})\Big)\bigg). \ene}
\begin{footnotesize}
   \bee\label{equ_tildaeta}\begin{split} 2\alpha(N_i\bE C_h-N\bE C_h^r+N_j\frac{S}{\lambda+S}\bE C_h^r-N_0\frac{\lambda_0}{\lambda_0+S}\bE C_h^r)
-2\alpha N_j(\frac{\lambda}{S}-\frac{\lambda}{\lambda+S})\bE C_h^r(\frac{1}{\kappa_j})^{\frac{\lambda+S}{\lambda-\lambda_0}}
-\alpha N_0(\frac{\lambda_0}{S}-\frac{\lambda_0}{\lambda_0+S})\bE C_h^r(\frac{1}{\kappa_j})^{\frac{\lambda_0+S}{\lambda-\lambda_0}}=0. \end{split}\ene\end{footnotesize}
\setcounter{equation}{\value{myver}}
\hrulefill
\end{figure*}

According to Proposition \ref{pro_bestrollovertime}, WSP $i$ will postpone its rollover time if the heavy user's overage cost reduction (from $\bE C_h$ in (\ref{equ_ECh}) to $\bE C_h^r$ in (\ref{equ_EChr})) is large and the new user population is small.
In reality, the heavy user's cost reduction after rollover is found not huge \cite{rollovernotsogood}, and we consider an upper bound for the cost reduction, i.e., $\bE C_h<\bE C_h^r\frac{2\lambda+S}{S}$.\footnote{For the case when $\bE C_h\geq\bE C_h^r\frac{2\lambda+S}{S}$, we can show that the equilibrium analysis is similar, and in the equilibrium results that only the medium $\eta_0$ regime includes more tedious cases to quantify equilibrium timing.}



\begin{figure} \centering
\subfigure[large $\eta_0$ regime or $\bE C_h\leq\bE C_h^r\frac{\lambda+S}{S}$ case] {\label{large}
\includegraphics[width=0.8\columnwidth]{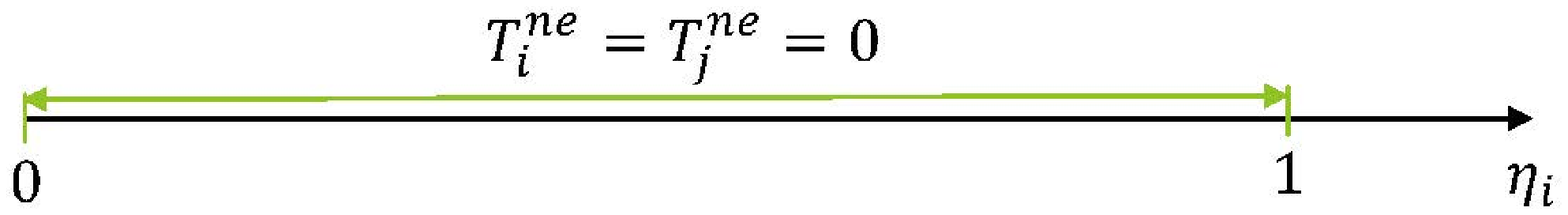}
}\\
\subfigure[Medium $\eta_0$ regime] {\label{medium}
\includegraphics[width=0.8\columnwidth]{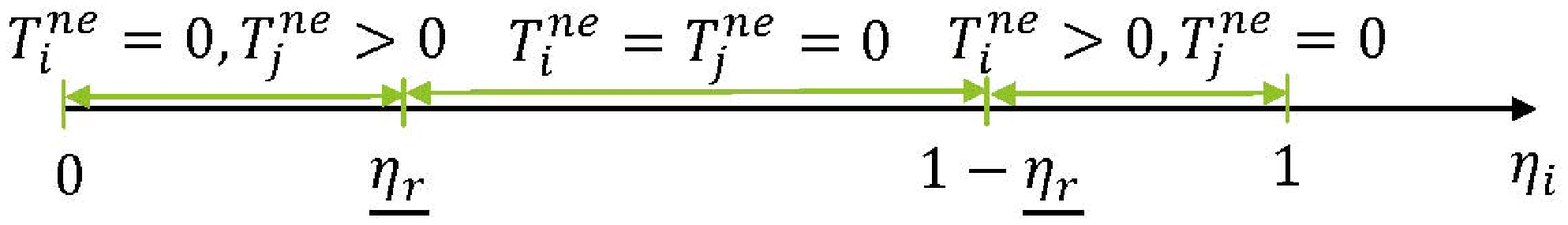}
}\\
\subfigure[Small $\eta_0$ regime] {\label{small}
\includegraphics[width=0.8\columnwidth]{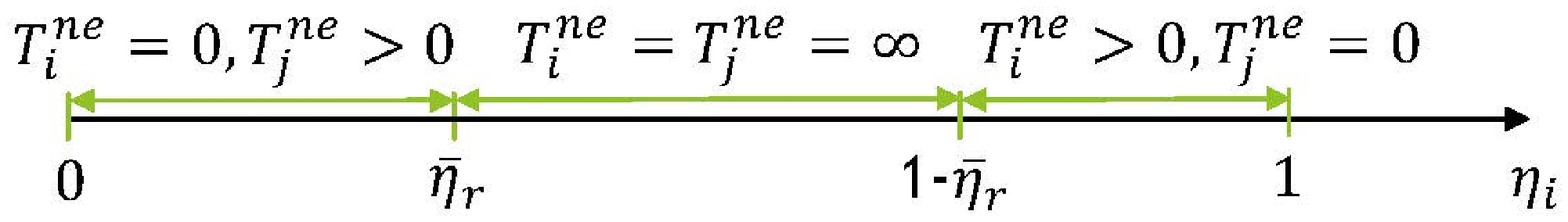}
}
\caption{Equilibrium rollover time vs new user proportion $\eta_0$ and WSP $i$'s market share.}
\label{NEvseta}
\end{figure}

\begin{thm}\label{thm_NE} The WSPs' equilibrium timing of rollover data plans is summarized as follows (see Fig. \ref{NEvseta}):\\
\begin{itemize}
  \item If $\bE C_h\leq\bE C_h^r\frac{\lambda+S}{S}$, the overage charge reduction after rollover is mild and the two WSPs will immediately upgrade their data plans (i.e., $T_i^{ne}=T_j^{ne}=0$), as shown in Fig. \ref{large}.
  \item If $\bE C_h>\bE C_h^r\frac{\lambda+S}{S}$, the WSPs' equilibrium rollover timing depends on the new user proportion $\eta_0$ and the WSPs' market shares:
      \begin{enumerate}
        \item \emph{Large $\eta_0$ regime} ($\eta_0\geq \frac{2S}{\lambda_0}(\frac{\bE C_h}{\bE C_h^r}-\frac{\lambda+S}{S})$): both WSPs will offer rollover data plans immediately to attract many new (heavy) users (i.e., $T_i^{ne}=T_j^{ne}=0$), as shown in Fig. \ref{large}.
        \item \emph{Medium $\eta_0$ regime} ($\bar{\eta}_0<\eta_0<\frac{2S}{\lambda_0}(\frac{\bE C_h}{\bE C_h^r}-\frac{\lambda+S}{S})$ with $\bar{\eta}_0$ as the solution to (\ref{equ_phi0})): the median size of new users and the other WSP's many users allure the WSP with a small market share (less than $\underline{\eta}_r$ in (\ref{equ_biglower})\footnote{$\tilde{\eta}_r$ in (\ref{equ_biglower}) is the solution of $R_i^{r,\leq}(T_i=0, T_j=\frac{\log\kappa_j}{\lambda-\lambda_0})=R_i^{r,>}(T_j=0, T_i=\frac{\log\kappa_i}{\lambda-\lambda_0})$ as a function of $\eta_i$.}) to upgrade immediately, and the other WSP purposely waits and upgrades at time $\frac{\log\kappa_i}{\lambda-\lambda_0}>0, i\in\{1,2\}$. If the two WSPs have similar market shares (between $\underline{\eta_r}$ and $1-\underline{\eta_r}$), they will upgrade immediately to keep their existing users and attract new users (see Fig. \ref{medium}).


      \item \emph{Small $\eta_0$ regime} ($0\leq\eta_0\leq\bar{\eta}_0$): the small size of new users cannot allure the two WSPs with similar market shares (between $\bar{\eta}_r$ and $1-\bar{\eta}_r$ with $\bar{\eta}_r$ as the solution of (\ref{equ_tildaeta})) to upgrade to rollover data plan (i.e., $T_i^{ne}=T_j^{ne}=\infty$). Only the WSP of small market share (less than $\bar{\eta}_r$) still chooses to upgrade to attract the other WSP's large market share, and the other will still upgrade after $\frac{\log\kappa_i}{\lambda-\lambda_0}, i\in\{1,2\}$ to keep its market share to some extent (see Fig. \ref{small}).
      \end{enumerate}
\end{itemize}
\newcounter{mytempeqncnt3}
\setcounter{mytempeqncnt3}{\value{equation}}
\setcounter{equation}{32}
\end{thm}


The proof of Theorem \ref{thm_NE} is given in Appendix A of the supplemental material. Then, we can conclude that
\begin{itemize}
  \item Both WSPs will offer rollover data plans immediately if the heavy user's cost reduction (i.e., $\bE C_h-\bE C_h^r$) is small or the new user population is large.
  \item The WSP with small market share prefers to rollover first as it can attract many users at least from the other WSP. While the WSP with large market share prefers late rollover to avoid the immediate revenue loss, and it still chooses rollover upgrade to keep its market share.
  \item Given few new users and large cost reduction from heavy users, the WSPs with similar market shares will not rollover as the profit received from the new users cannot compensate for the revenue loss due to users' cost reduction.
\end{itemize}

These results match well with our industry observations about rollover data plans. For example, in China, the quarterly new customers growth is larger than 10\%, which falls into the large new user regime in Fig. \ref{large}. Thus the two major WSPs, China Mobile and China Unicom, announced the rollover data plan at the same time \cite{Chinarollorder}. In the USA, the quarterly new customers growth is around 2\% and the WSPs with diverse market shares chose rollover sequentially, i.e., T-mobile with the small market share announced rollover first in 2014, which is followed by AT\&T in 2015 \cite{USrollover}. In a saturated market like Singapore, the quarterly new customers growth is trivial, which can be viewed as the small new user regime. So far, no WSP in this market has announced rollover data since the involved WSPs' market shares are similar.

\begin{figure*}[t]
\centering
\subfigure[Large $\eta_0$ regime]{\label{largeprofit}
\begin{minipage}{.315\textwidth}
\includegraphics[width=1\textwidth]{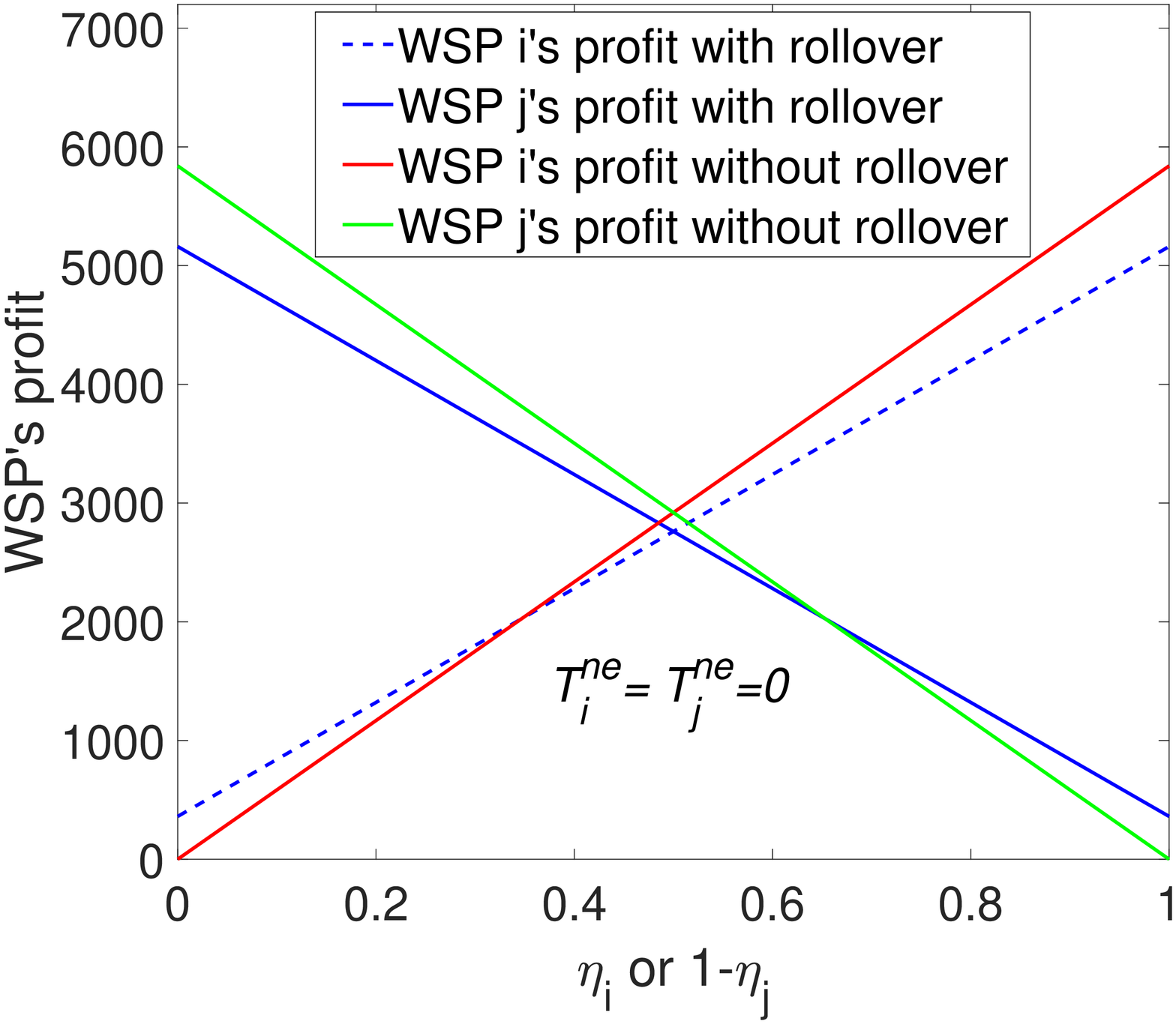}
\end{minipage}
}
\subfigure[Medium $\eta_0$ regime]{\label{mediumprofit}
\begin{minipage}{.315\textwidth}
\includegraphics[width=1\textwidth]{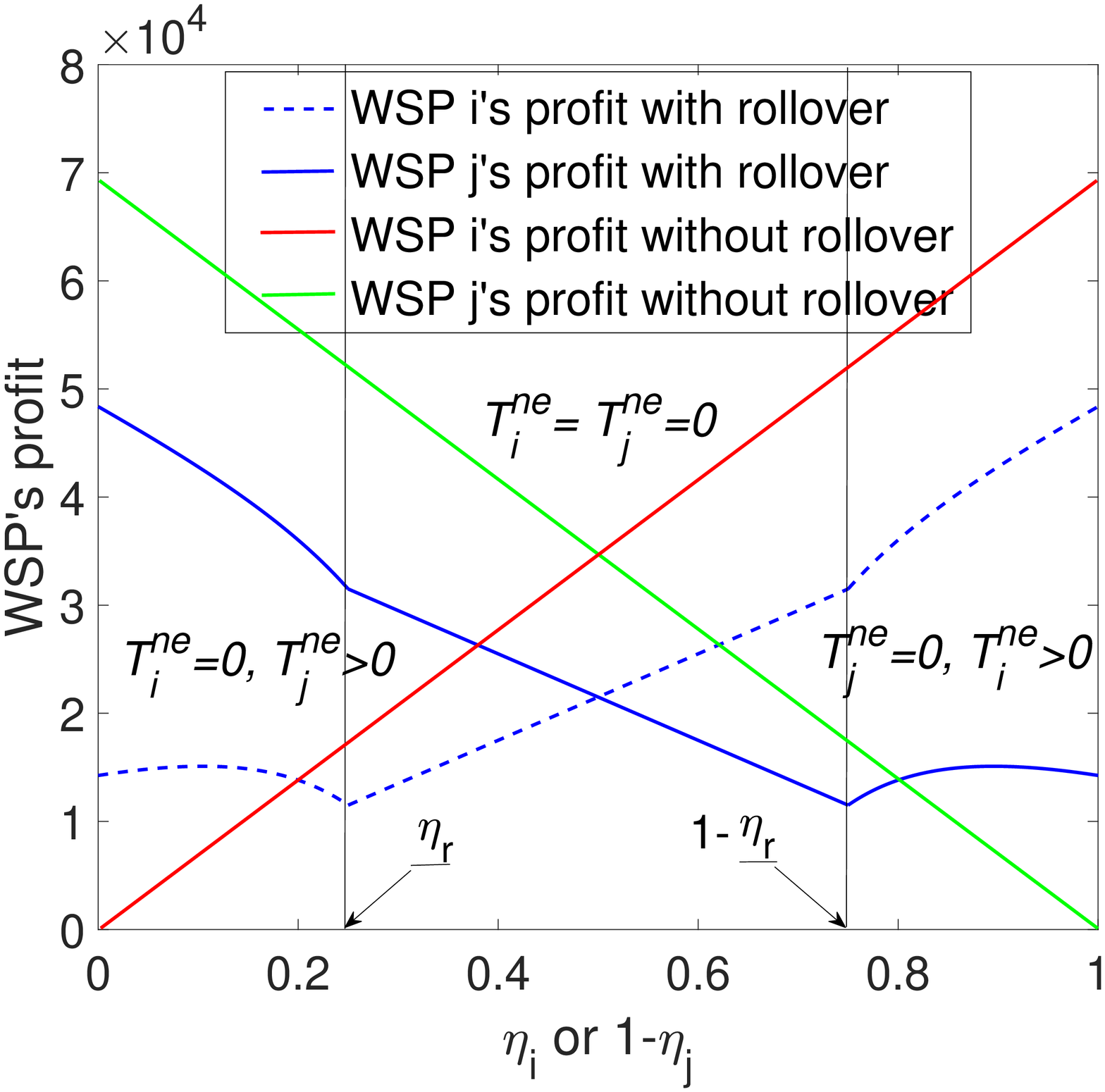}
\end{minipage}
}
\subfigure[Small $\eta_0$ regime]{\label{smallprofit}
\begin{minipage}{.32\textwidth}
\includegraphics[width=1\textwidth]{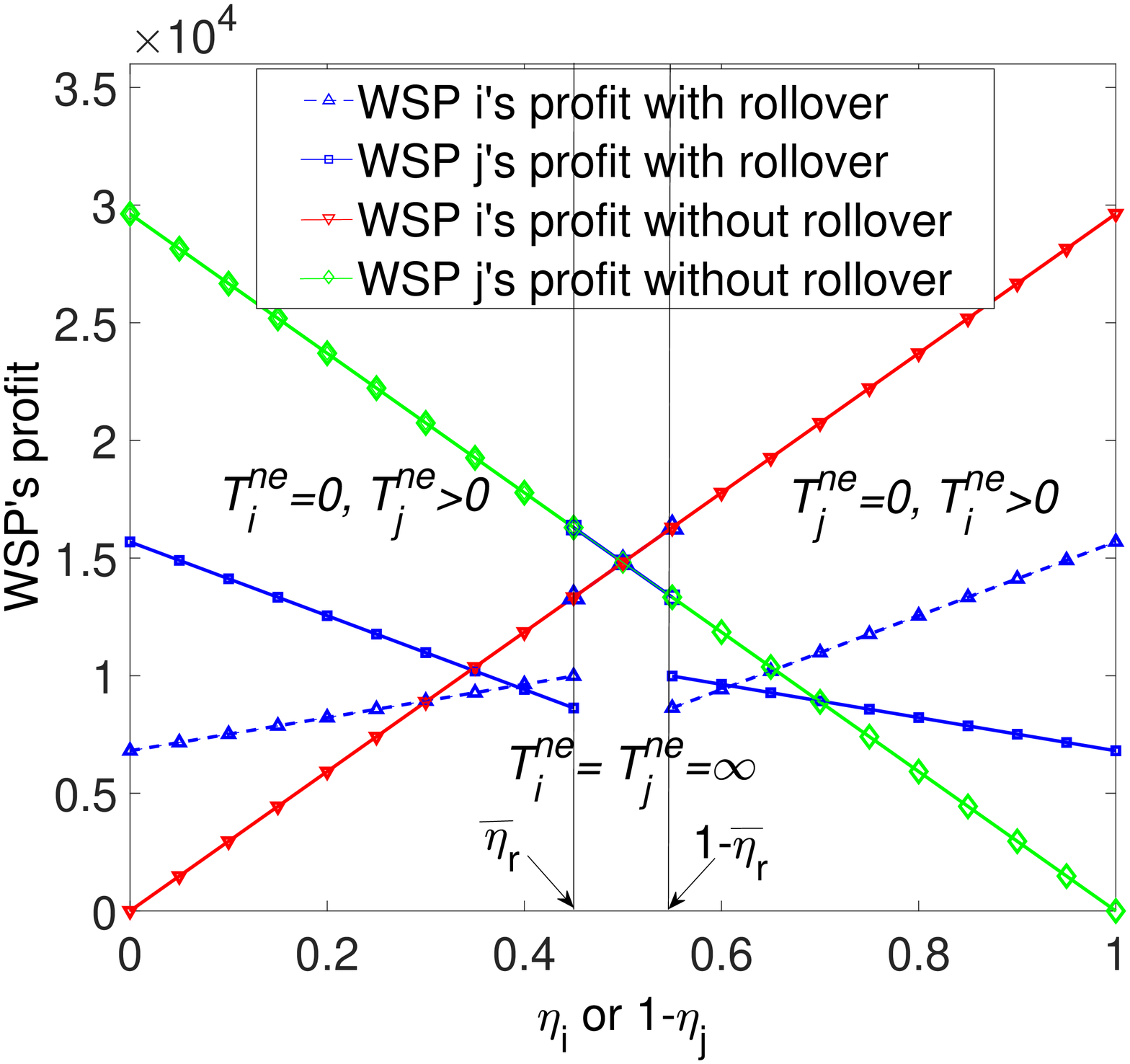}
\end{minipage}
}
\caption{WSPs' equilibrium profits at equilibrium rollover time versus WSP $i$'s market share ($\eta_i$ or $1-\eta_j$) and new user proportion $\eta_0$.}\label{NEprofit}
\end{figure*}

\subsection{Multiple WSPs' Equilibrium Timing}

\begin{table}[htbp]
\centering
\caption{Equilibrium Timing for three WSPs with market shares $\eta_1=0.1, \eta_2=0.3, \eta_3=0.6$.
The expected cost $\bE C_h^r$ under rollover data plan is fixed as $3$, and we change the new user population $\eta_0$ and the initial cost $\bE C_h$ to alter $\Delta=\bE C_h-\bE C_h^r$.}\label{table_NE_multi_WSPs}
\begin{tabular}{|l|l|l|}
\hline \textbf{$\bE C_h$} & \textbf{$\eta_0$} & \textbf{Equilibrium Timing}
\\\hline $(3,5.6]$ & $0.3$ & $T_1^{ne}=T_2^{ne}=T_3^{ne}=0$
\\\hline $(5.6,8.5]$ & $0.3$ & $0=T_1^{ne}=T_2^{ne}<T_3^{ne}$
\\\hline $(8.5,10]$ & $0.3$ & $0=T_1^{ne}<T_2^{ne}<T_3^{ne}$
\\\hline $(10,18]$ & $0.3$ & $0=T_1^{ne}<T_2^{ne}=T_3^{ne}$
\\\hline $(18,\infty]$ & $0.3$ & $T_1^{ne}=T_2^{ne}=T_3^{ne}=\infty$
\\\hline $5.5$ & $0.1$ & $0=T_1^{ne}=T_2^{ne}<T_3^{ne}$
\\\hline
\end{tabular}
\end{table}

For multiple asymmetric WSPs, we can numerically show the equilibrium rollover timing in Table \ref{table_NE_multi_WSPs}, where the three WSPs have market shares $\eta_1=0.1, \eta_2=0.3, \eta_3=0.6$ initially. As shown in Table \ref{table_NE_multi_WSPs}, when the overage charge cost reduction after rollover is mild, i.e., from initial cost $\bE C_h\in(3,5.6]$ to $\bE C_h^r=3$ under the rollover data plan, the three WSPs will immediately
upgrade their data plans, i.e., $T_1^{ne}=T_2^{ne}=T_3^{ne}=0$ to equally attract the sizable new users with large churn rate $\eta_0=0.3$.
As the cost reduction increases, i.e., $\bE C_h\in(5.6,8.5]$, the WSPs
with smaller market share will upgrade immediately to attract the new users and other WSP's large market share, i.e., $T_1^{ne}=T_2^{ne}=0$,
and the WSP with the largest market share purposely waits and upgrades at time $T_3^{ne}>0$ to avoid the immediate revenue
loss from its existing users. If the cost reduction further increases, i.e., $\bE C_h\in(8.5,10]$,
only the WSP with the smallest market share will upgrade immediately and the other two WSPs will wait and upgrade successively according to
their current market share, i.e., $0<T_2^{ne}<T_3^{ne}$. For larger cost reduction regime $\bE C_h\in(10,18]$, only the WSP with the smallest
market share will upgrade immediately, and the second largest WSP will further postpone its upgrade timing and upgrade together with the largest WSP
$3$, i.e., $T_2^{ne}=T_3^{ne}>0$. For very large cost reduction regime $\bE C_h\in(18,\infty]$, the WSPs will not upgrade to
rollover data plan, i.e., $T_1^{ne}=T_2^{ne}=T_3^{ne}=\infty$, as the increased profit from the switched users and new users cannot
compensate for the profit loss from existing users. Moreover, we show that for the mild cost reduction, i.e., $\bE C_h=5.5$, as the proportion
of new users $\eta_0$ decreases from $0.3$ to $0.1$, the equilibrium rollover timing will change from $T_1^{ne}=T_2^{ne}=T_3^{ne}=0$ to
$0=T_1^{ne}=T_2^{ne}<T_3^{ne}$. That is to say, the WSP $3$ with the largest market share will postpone its upgrade time as the small new user
population cannot justify the reduction of immediate overage charges due to rollover. The numerical results shown in Table \ref{table_NE_multi_WSPs}
for multiple WSPs coincide with the key results of two-WSP case in Theorem \ref{thm_NE} and Fig. \ref{NEvseta}. Moreover, for multiple symmetric WSPs, we have the
following corollary, which is consistent with Theorem \ref{thm_NE}.

\begin{cor}\label{cor_moreWSPs} For multiple symmetric WSPs, they will offer rollover immediately at the same time if $\bE C_h-\bE C_h^r\frac{\lambda+S}{S}<\eta_0\bE C_h^r\frac{\lambda_0}{S}$ for large $\eta_0$ or mild cost reduction.
\end{cor}

\textbf{Proof:} Suppose there are $\cM\geq 2$ symmetric WSPs. Note that a WSP's existing users are only affected by the WSP who offers rollover first as the existing users start to churn to the earlier rollover WSP gradually. For any WSP $i\in\cM$, denote $T_0$ as the WSPs who offers rollover first. 
Then, WSP $i\in\cM$ will not offer rollover latest if \bee\begin{split} \frac{\partial R_i^{r,>}}{\partial T_i}=&2\alpha \frac{N}{\cM}(\bE C_h-\bE C_h^r\frac{\lambda+S}{S})e^{\lambda T_0-(\lambda+S)T_i}\\
&-2\alpha \frac{N_0}{\cM}\bE C_h^r\frac{\lambda_0}{S}e^{\lambda_0 T_0-(\lambda_0+S)T_i}<0, \end{split}\ene
i.e., $\frac{N}{\cM}(\bE C_h-\bE C_h^r\frac{\lambda+S}{S})<\frac{N_0}{\cM}\bE C_h^r\frac{\lambda_0}{S}$. As a result, all WSPs prefer rollover earlier, which results in the equilibrium point that all WSPs offer rollover at the beginning. \qed

\begin{rem} Note that for multiple symmetric WSPs, they may have asymmetric upgrade timing. For example, for non-trivial overage charge cost reduction and not large new user population, e.g., $\bE C_h=6, \bE C_h^r=3, \eta_0=0.3$, two out of the three WSPs with market shares $\eta_1=\eta_2=\eta_3$ will upgrade immediately to attract the new users and switched users,
i.e., $0=T_1^{ne}=T_2^{ne}<T_3^{ne}$. Given the equilibrium rollover timing of the two WSPs, if the third WSP still chooses to upgrade at the beginning,
it can only attract the new users and no switched users from the other WSPs will join it. Therefore, it is better for the third WSP to
wait and upgrade at time $T_3^{ne}>0$ as the revenue received from the new users cannot compensate for the revenue loss from existing users.
\end{rem}

\subsection{WSPs' Profits at Equilibrium Timing}

Now we are ready to discuss the duopoly WSPs' profits at the equilibrium. By substituting the equilibrium rollover time in Theorem \ref{thm_NE} to (\ref{equ_R_i^r_early}) and (\ref{equ_R_i^r_late}), we can derive the equilibrium profits of the two WSPs. At the equilibrium rollover timing, the profit of WSP $i, i=1,2$ is given as follows:
            \begin{itemize}
              \item The WSP $i$'s profit for rollover at the same time (i.e., $T_i^{ne}=T_j^{ne}=0$) is
              \bee\label{equ_profitequal} \bar{R}_i^r=2\alpha N_i\bE C_h^r\frac{1}{S}+\alpha N_0(\frac{1}{S}-\frac{1}{S+\lambda_0})\bE C_h^r. \ene
              \item The WSP $i$'s profit for first rollover (i.e., $T_i^{ne}=0$, $T_j^{ne}=\frac{\log\kappa_j}{\lambda-\lambda_0}$) is
\bee\label{equ_profitearly}\begin{split} \bar{R}_i^{r,\leq}=&2\alpha N\bE C_h^r\frac{1}{S}-2\alpha N_j\bE C_h^r\frac{1}{S+\lambda}\\
&-2\alpha N_j\bE C_h^r(\frac{1}{S}-\frac{1}{\lambda+S})(\frac{1}{\kappa_j})^{\frac{\lambda+S}{\lambda-\lambda_0}}\\
&+2\alpha N_0(\frac{1}{S}-\frac{1}{S+\lambda_0})\bE C_h^r\\
&-\alpha N_0(\frac{1}{S}-\frac{1}{S+\lambda_0})(\frac{1}{\kappa_j})^{\frac{\lambda_0+S}{\lambda-\lambda_0}}\bE C_h^r. \end{split}\ene
              \item The WSP $i$'s profit for late rollover (i.e., $T_i^{ne}=\frac{\log\kappa_i}{\lambda-\lambda_0}$, $T_j^{ne}=0$) is
\bee\label{equ_profitlate}\begin{split} \bar{R}_i^{r,>}=&2\alpha N_i\frac{1}{\lambda+S}\bE C_h(1-(\frac{1}{\kappa_i})^{\frac{\lambda+S}{\lambda-\lambda_0}})\\
&+2\alpha N_i\frac{1}{S}\bE C_h^r(\frac{1}{\kappa_i})^{\frac{\lambda+S}{\lambda-\lambda_0}}\\
&+\alpha N_0(\frac{1}{S}-\frac{1}{S+\lambda_0})(\frac{1}{\kappa_i})^{\frac{\lambda_0+S}{\lambda-\lambda_0}}\bE C_h^r. \end{split}\ene
            \end{itemize}

If the new user population is extremely large, it is manifest that the WSPs always gain profit by upgrading to rollover data. Next, we look at the non-trivial case that the number of new users is smaller than that of the existing users, i.e., $\eta_0<1$. By comparing WSP $i$'s profits before and after rollover at the equilibrium, we have the following proposition.

\begin{pro}\label{pro_rollover_threshold} At the equilibrium, there exists a unique threshold $\eta_i^{th}\in [0,1]$ such that WSP $i$ gains profit by data rollover if $\eta_i\leq \eta_i^{th}$, and loses profit if $\eta_i>\eta_i^{th}$.
\end{pro}

The Proof of Proposition \ref{pro_rollover_threshold} is given in Appendix B of the supplemental material.

\newcounter{my6}
\begin{figure*}[ht]
\setcounter{my6}{\value{equation}}
\setcounter{equation}{36}
\footnotesize{
\bee\label{equ_phiphi}\begin{split} &\frac{1}{2}\cD(\frac{\lambda}{\lambda+S}+\frac{1}{2}\frac{S}{\lambda+S})-\cE(\frac{\lambda}{\lambda+S}+\frac{1}{4}\frac{S}{\lambda+S})
-\cE\eta_0\frac{\lambda_0}{\lambda_0+S}-\frac{1}{2}\cE(\frac{\lambda}{S}-\frac{\lambda}{\lambda+S})\Big(\frac{\cE\eta_0\frac{\lambda_0}{S}}{\frac{1}{2}(\cD-\cE)-\cE\frac{\lambda}{S}}\Big)^{\frac{\lambda+S}{\lambda-\lambda_0}}\\
&-\frac{1}{2}\cE\eta_0\frac{\lambda_0^2}{S(\lambda_0+S)}\Big(\frac{\cE\eta_0\frac{\lambda_0}{S}}{\frac{1}{2}(\cD-\cE)-\cE\frac{\lambda}{S}}\Big)^{\frac{\lambda_0+S}{\lambda-\lambda_0}}=0. \end{split}\ene
}
\footnotesize{
\bee\label{equ_underlineeta_s}\begin{split} \underline{\eta_s}=\max\Big(1-\tilde{\eta}_s, \min\big(\max(1-\hat{\eta}_s, \frac{\cE\frac{\lambda}{S}+\sqrt{(\cE\frac{\lambda}{S})^2+2(\cD-\cE)\cE\eta_0\frac{\lambda_0}{S}}}{2(\cD-\cE)}), \min(\hat{\eta}_s, 1-\frac{\cE\frac{\lambda}{S}+\sqrt{(\cE\frac{\lambda}{S})^2+2(\cD-\cE)\cE\eta_0\frac{\lambda_0}{S}}}{2(\cD-\cE)})\big)\Big). \end{split}\ene
\bee\label{equ_veta}\begin{split} &\cD\eta_i(\frac{\lambda}{\lambda+S}+\eta_i\frac{S}{\lambda+S})-\cE(\frac{\lambda}{\lambda+S}+\eta_i^2\frac{S}{\lambda+S})
-\cE\eta_0\frac{\lambda_0}{\lambda_0+S}-\eta_j\cE(\frac{\lambda}{S}-\frac{\lambda}{\lambda+S})(\frac{1}{\kappa_j^s})^{\frac{\lambda+S}{\lambda-\lambda_0}}
-\frac{1}{2}\cE\eta_0\frac{\lambda_0^2}{S(\lambda_0+S)}(\frac{1}{\kappa_j^s})^{\frac{\lambda_0+S}{\lambda-\lambda_0}}=0. \end{split}\ene}
\setcounter{equation}{\value{my6}}
\hrulefill
\end{figure*}

\begin{figure}
\centering\includegraphics[scale=0.45]{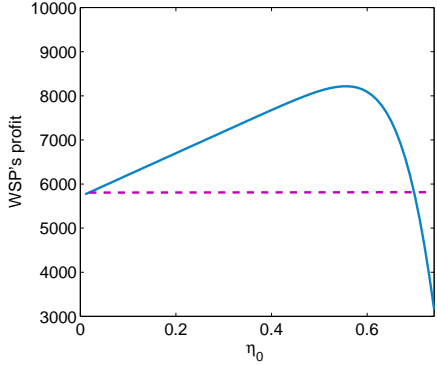}\caption{WSP's equilibrium profit (the blue solid curve) under first rollover versus $\eta_0$.}\label{WSPprofitrollfirst}
\end{figure}

If a WSP has a small market share, it can still increase its profit by first rollover and attracting many users from the other WSP. While for the WSP with already a large market share, the small new user population cannot justify the reduction of overage charges due to rollover. In the following, we present numerical results in Figs. \ref{NEprofit} and \ref{WSPprofitrollfirst} to examine the WSPs's equilibrium profits.

\emph{Observation 1:} A WSP's equilibrium profit can decrease with its market share.

As shown in Fig.\ref{largeprofit} and \ref{mediumprofit}, if both WSPs choose rollover immediately, it is manifest that each WSP's profit increases with its market share. However, in Fig. \ref{mediumprofit}, we observe that if WSP $i$ provides rollover data plan first, its profit for first rollover may decrease with its market share $\eta_i$. This is because as $\eta_i$ increases, WSP $j$'s market share $\eta_j$ decreases and it brings forward its rollover time, which results in the number of new users and the existing users switched from WSP $j$ becomes smaller. Similarly, in Fig. \ref{smallprofit}, the WSP $i$'s profit suddenly reduces as $\eta_i$ increases across the threshold point $1-\bar{\eta}_r$ (i.e., from medium to large market share). The reason is that if $\eta_i$ is slightly larger than $1-\bar{\eta}_r$, WSP $i$ prefers rollover late and the revenue received from the few new users cannot compensate for the revenue loss due to its many existing users' cost reduction. 

\emph{Observation 2:} The WSP may not benefit from the new user population.

As shown in Fig. \ref{WSPprofitrollfirst}, the WSP's profit for earlier rollover first increases with $\eta_0$, as it gains more revenue from more new users. However, as $\eta_0$ further increases, the WSP's profit for earlier rollover decreases. This is because the other WSP's rollover time is brought forward (with intensified competition), which reduces the number of new users and switched existing users  subscribed to the WSP of earlier rollover.



\section{Timing Equilibrium for Shared Data Plan}\label{sec_share}

When deciding the timing for shared data plan, a WSP also faces the tradeoff between its overage charge loss from its existing users and the profit gain of market share (from new users and those from the other WSP). Recall that in the earlier upgrade profit in (\ref{equ_R_se}), we have introduced $\cD$ and $\cE$ in (\ref{equ_cD}) and (\ref{equ_cE}) to tell the families' expected costs before and after using the shared data plan, respectively. We note that $\cD>\cE$ as the (heavy, heavy) and (heavy, light) families' costs are reduced. 
One can imagine that if the cost reduction after offering the shared data plan is huge, the WSPs prefer not to offer due to large revenue loss. In the following analysis, we consider an upper bound for such cost reduction, i.e., $\cD<\cE\frac{4\lambda+S}{S}$.\footnote{For the case when $\cD\geq\cE\frac{4\lambda+S}{S}$, we can show that the equilibrium analysis is similar, and in the equilibrium results that only the medium $\eta_0$ regime includes more tedious cases to quantify equilibrium timing.} Similar to the timing analysis of the rollover data plan, we first analyze the WSPs' best responses and then derive the WSPs' equilibrium timing of shared data plan as follows. 


\begin{figure} \centering
\subfigure[Large $\eta_0$ regime or $\cD\leq\cE\frac{\lambda+S}{S}$ case] { \label{largeshare}
\includegraphics[width=0.8\columnwidth]{largeeta0}
}\\
\subfigure[Medium $\eta_0$ regime] { \label{mediumshare}
\includegraphics[width=0.8\columnwidth]{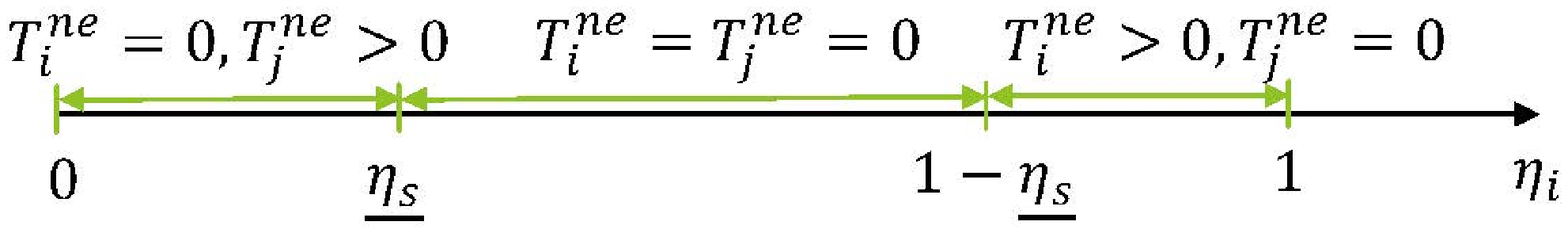}
}\\
\subfigure[Small $\eta_0$ regime (if exists)] { \label{smallshare}
\includegraphics[width=0.8\columnwidth]{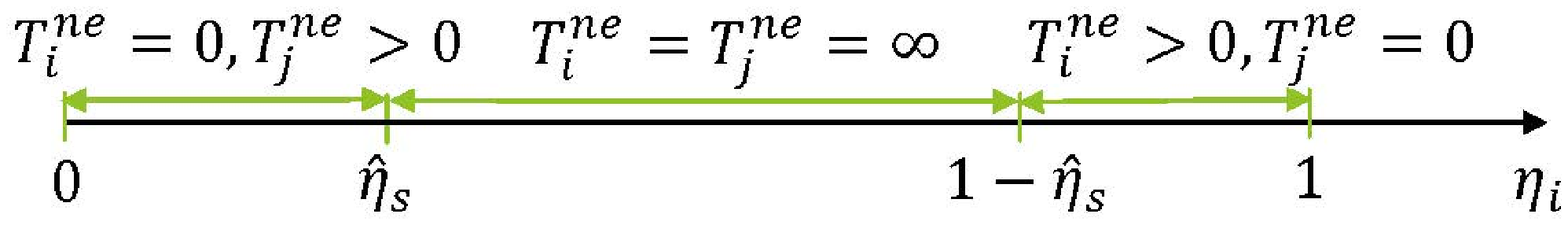}
}
\caption{Equilibrium timing vs new user population $\eta_0$ and WSP $i$'s market share.}
\label{NEvsetashare}
\end{figure}

\begin{figure*}[t]
\centering
\subfigure[Large $\eta_0$ regime]{\label{largeprofitshare}
\begin{minipage}{.315\textwidth}
\includegraphics[width=1\textwidth]{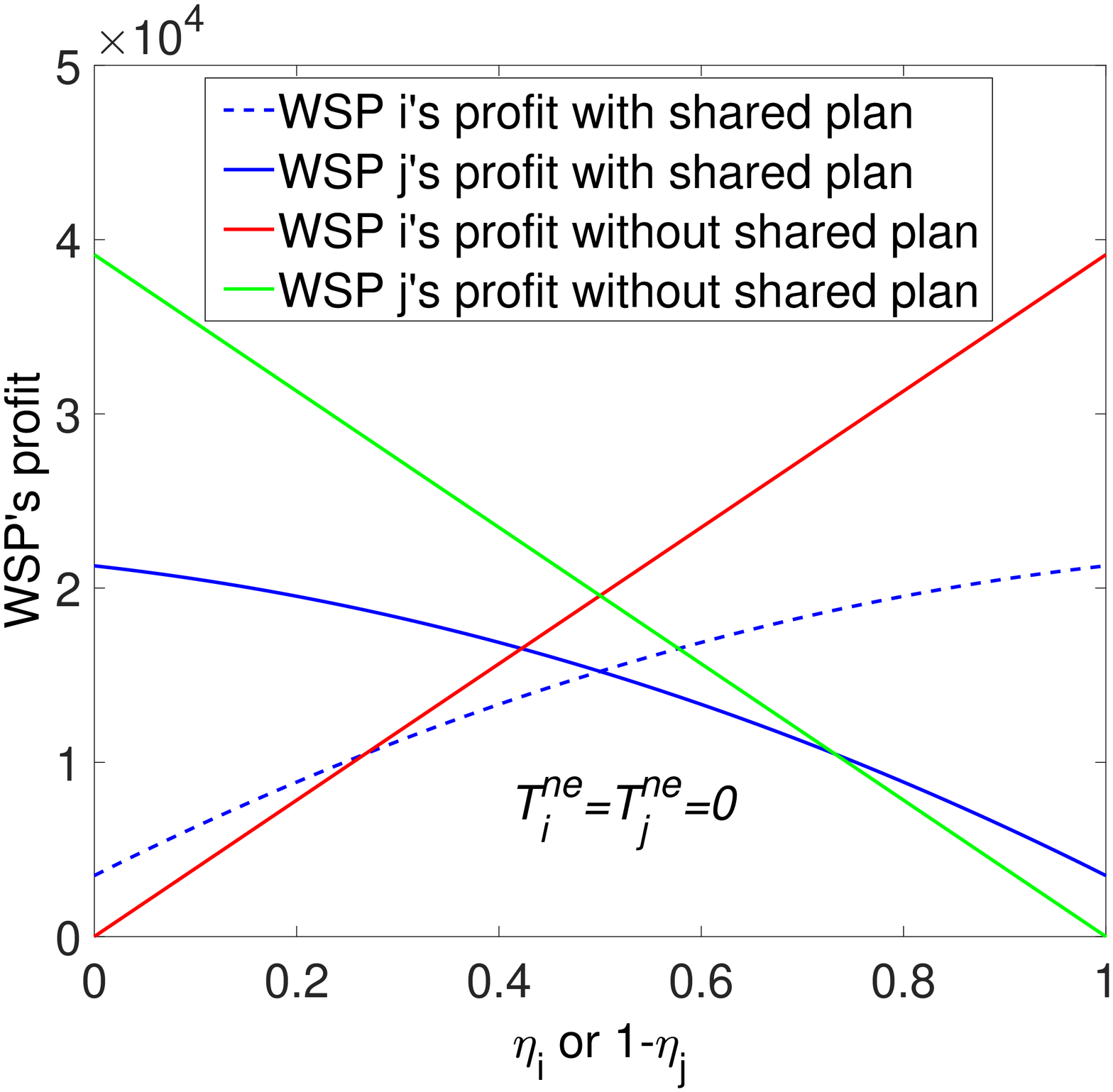}
\end{minipage}
}
\subfigure[Medium $\eta_0$ regime]{\label{mediumprofitshare}
\begin{minipage}{.315\textwidth}
\includegraphics[width=1\textwidth]{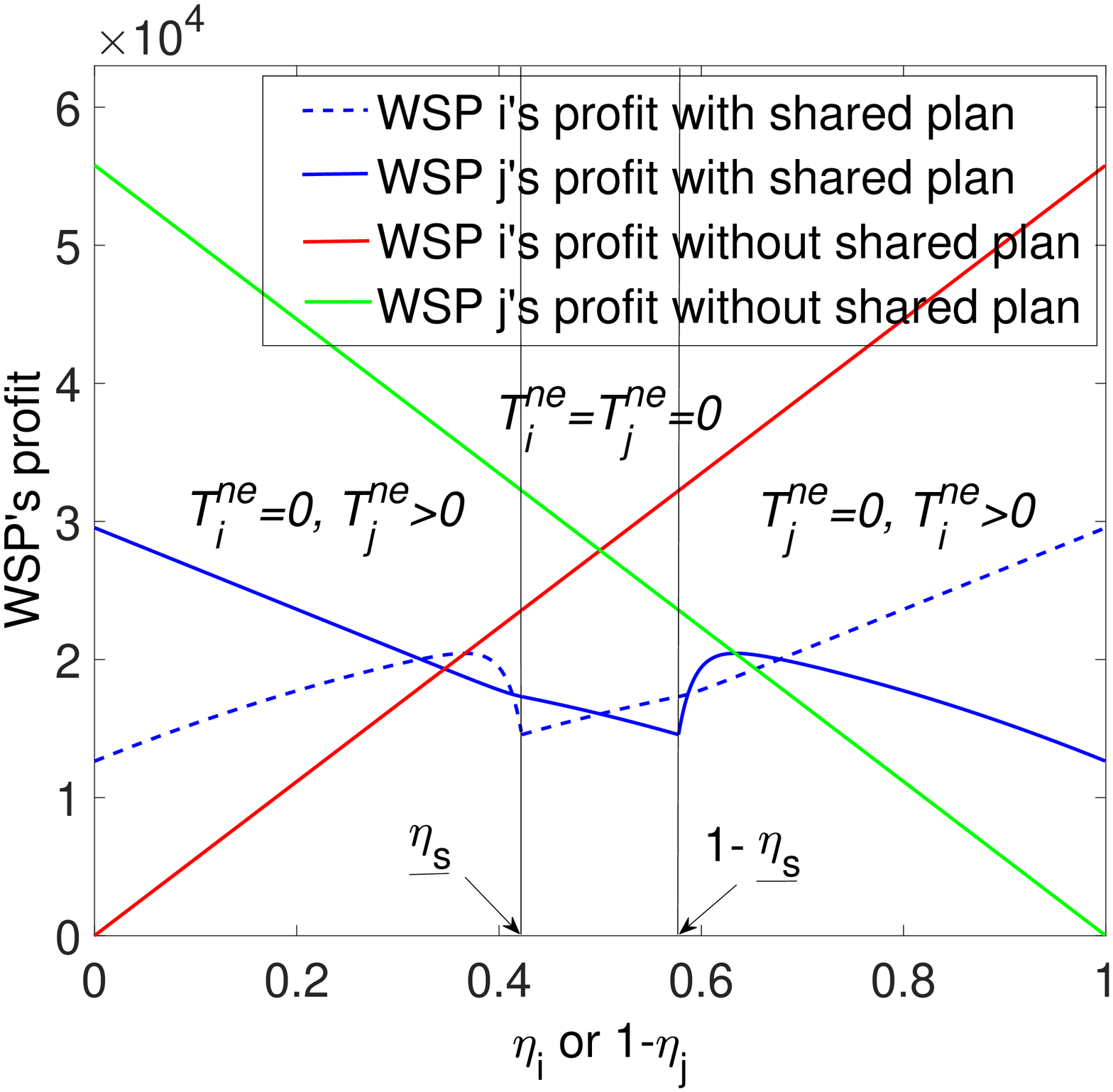}
\end{minipage}
}
\subfigure[Small $\eta_0$ regime]{\label{smallprofitshare}
\begin{minipage}{.316\textwidth}
\includegraphics[width=1\textwidth]{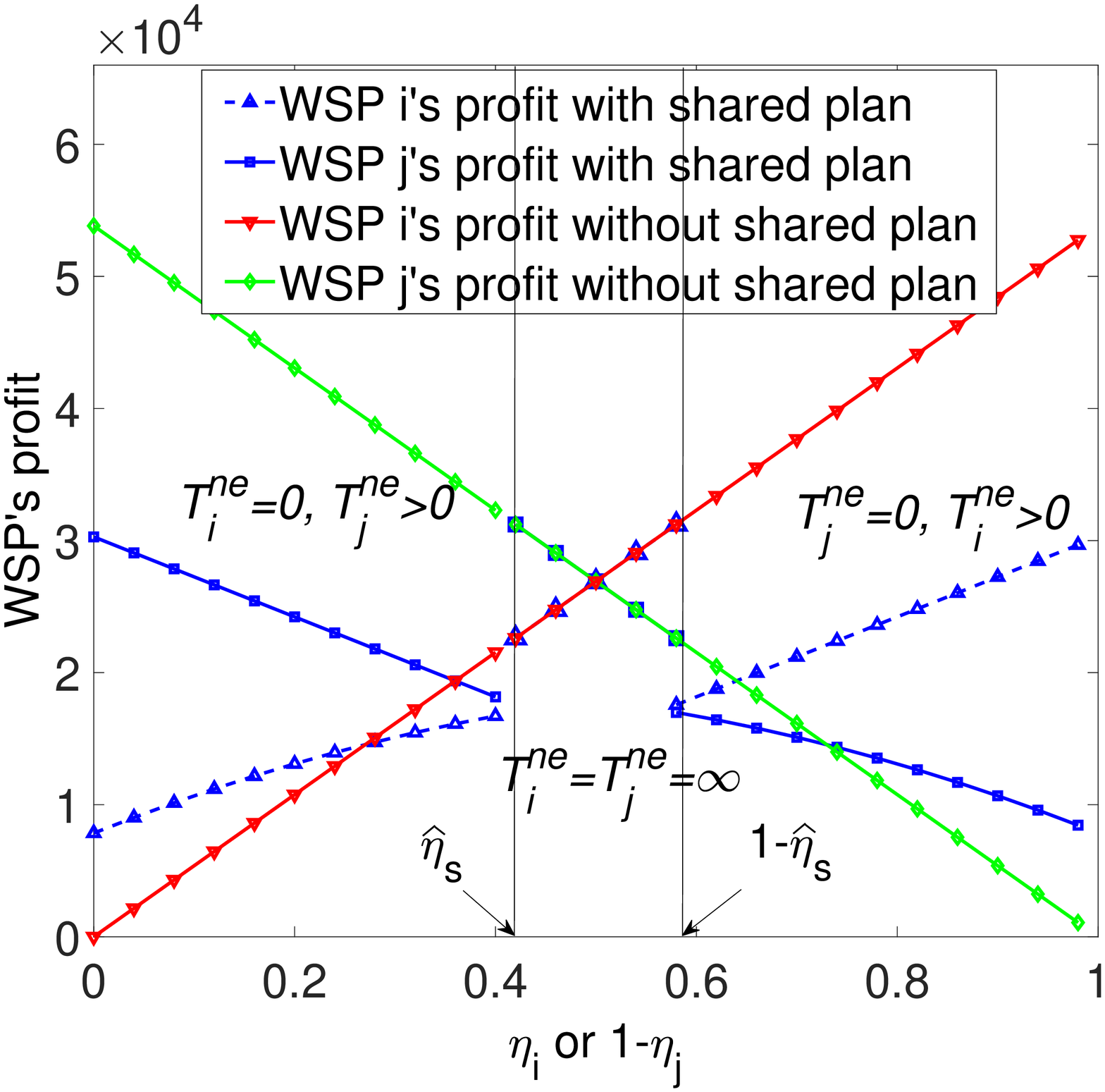}
\end{minipage}
}
\caption{WSPs' equilibrium profits at equilibrium share time versus WSP $i$'s market share ($\eta_i$ or $1-\eta_j$) and new user proportion $\eta_0$.}\label{NEprofitshare}
\end{figure*}


\newcounter{my7}
\setcounter{my7}{\value{equation}}
\setcounter{equation}{39}

\begin{thm}\label{thm_NEshare} The WSPs' equilibrium upgrade timing of shared data plan is given as follows (see Fig. \ref{NEvsetashare}):\\
\begin{itemize}
  \item If $\cD\leq\cE\frac{\lambda+S}{S}$, the overage charge reduction after offering shared data plan is mild and the two WSPs will immediately upgrade their data plans (i.e., $T_i^{ne}=T_j^{ne}=0$), as shown in Fig. \ref{largeshare}.
  \item If $\cD>\cE\frac{\lambda+S}{S}$, the WSPs' equilibrium share timing depends on the new user proportion $\eta_0$ and the WSPs' market shares:
  \begin{enumerate}
    \item \emph{Large $\eta_0$ regime} ($\eta_0\geq \frac{2(\cD S-\cE(\lambda+S))}{\cE\lambda_0}$): both WSPs will offer shared data plan immediately to attract many new (heavy, heavy) and (heavy, light) families (i.e., $T_i^{ne}=T_j^{ne}=0$), as shown in Fig. \ref{largeshare}.
  \item \emph{Medium $\eta_0$ regime} ($\min((\bar{\eta}_0^s)^+,(\frac{\cD S-\cE(2\lambda+S)}{2\cE\lambda_0})^+)<\eta_0<\frac{2(\cD S-\cE(\lambda+S))}{\cE\lambda_0}$ with $\bar{\eta}_0^s$ as the solution to (\ref{equ_phiphi})): the median size of new users and the other WSP's many users allure the WSP with a small market share (less than $\underline{\eta_s}$ in (\ref{equ_underlineeta_s})\footnote{$\tilde{\eta}_s$ in (\ref{equ_underlineeta_s}) is the solution of $R_i^{s,\leq}(T_i=0, T_j=\frac{\log\kappa_j^s}{\lambda-\lambda_0})=R_i^{s,>}(T_j=0, T_i=\frac{\log\kappa_i^s}{\lambda-\lambda_0})$ as a function of $\eta_i$ for any given $\eta_0$, with $R_i^{s,\leq}$ and $R_i^{s,>}$ given in (\ref{equ_R_se}) and (\ref{equ_R_sl}) respectively.}) to upgrade immediately, and the other WSP purposely waits for $\frac{\log\kappa_i^s}{\lambda-\lambda_0}, i\in\{1,2\}>0$, where \begin{equation*}\setlength{\abovedisplayskip}{3pt}
\setlength{\belowdisplayskip}{3pt} \kappa_i^s=\frac{(\cD-\cE)\eta_i^2-\cE\frac{\lambda}{S}\eta_i}{\frac{1}{2}\cE\eta_0\frac{\lambda_0}{S}}. \end{equation*}
If the two WSPs have similar market shares (between $\underline{\eta_s}$ and $1-\underline{\eta_s}$), they will upgrade immediately to keep their existing users and attract new users (see Fig. \ref{mediumshare}). 

  \item \emph{Small $\eta_0$ regime} ($0\leq\eta_0\leq\min((\bar{\eta}_0^s)^+,(\frac{\cD S-\cE(2\lambda+S)}{2\cE\lambda_0})^+)$): the small size of new users cannot allure the two WSPs with similar market shares (between $\hat{\eta}_s$ and $1-\hat{\eta}_s$ with $\hat{\eta}_s$ as the solution to (\ref{equ_veta})) to upgrade to shared data plan (i.e., $T_i^{ne}=T_j^{ne}=\infty$). Only the WSP of small market share (less than $\hat{\eta}_s$) still chooses to upgrade to attract the other WSP's large market share, and the other will still upgrade after $\frac{\log\kappa_i^s}{\lambda-\lambda_0}, i\in\{1,2\}$ to keep its market share to some extent (see Fig. \ref{smallshare}).
  \end{enumerate}
\end{itemize}


\end{thm}

The proof of Theorem \ref{thm_NEshare} is given in Appendix C of the supplemental material. Unlike Theorem \ref{thm_NE} for the rollover data plan, here the small $\eta_0$ regime in Fig. \ref{smallshare} does not exist if $\frac{\cD S-\cE(2\lambda+S)}{2\cE\lambda_0}\leq 0$. Then, we only have large and medium $\eta_0$ regimes in Fig. \ref{largeshare} and \ref{mediumshare}, where at least one WSP will upgrade to shared data plan even for few new user population. Recall when $\eta_0=0$ in the rollover data case, the WSPs can only benefit from the heavy users, and the WSPs with similar market share will not offer rollover as shown in Fig. \ref{small}. However, this is not true for the shared data case as a WSP benefits from attracting (heavy,heavy) and (heavy,light) families from the other WSP. Here, the relative proportion between (heavy, heavy) and (heavy, light) families increases with the proportion of heavy users $\alpha$. In the following proposition, we discuss this different result clearly given no new user arrival. 

\begin{pro}\label{pro_sharealways} Even if $\eta_0=0$, both WSPs with similar market shares (between $\underline{\eta_s}$ and $1-\underline{\eta_s}$ in Fig. \ref{mediumshare}) will still offer the shared data plan, once one of the following three conditions is true:
\begin{itemize}
  \item \emph{Large (heavy, heavy) family proportion with mild cost reduction:} If (heavy, light) family's cost reduction is significant (i.e., $\frac{\bE C_h+\bE C_l}{\bE C_{h,l}}>\frac{2\lambda+S}{S}$) but (heavy, heavy) family's cost reduction is mild (i.e., $\frac{2\bE C_h}{\bE C_{h,h}}\leq\frac{2\lambda+S}{S}$), then the WSPs will offer shared data plan when facing large proportion of (heavy, heavy) families (i.e., $\alpha\geq\frac{2(\bE C_h+\bE C_l-\frac{2\lambda+S}{S}\bE C_{h,l})}{2\bE C_l+(\bE C_{h,h}-2\bE C_{h,l})\frac{2\lambda+S}{S}}$).
  \item \emph{Large (heavy, light) family proportion with mild cost reduction:} If (heavy, light) family's cost reduction is mild (i.e., $\frac{\bE C_h+\bE C_l}{\bE C_{h,l}}\leq\frac{2\lambda+S}{S}$) but (heavy, heavy) family's cost reduction is significant (i.e., $\frac{2\bE C_h}{\bE C_{h,h}}>\frac{2\lambda+S}{S}$), then the WSPs will offer shared data plan when facing small proportion of (heavy, heavy) families (i.e., $\alpha\leq\frac{2(\bE C_h+\bE C_l-\frac{2\lambda+S}{S}\bE C_{h,l})}{2\bE C_l+(\bE C_{h,h}-2\bE C_{h,l})\frac{2\lambda+S}{S}}$).
  \item \emph{Mild cost reduction for both (heavy, heavy) and (heavy, light) families:} If both (heavy, heavy) and (heavy, light) families' cost reductions are mild (i.e., $\frac{\bE C_h+\bE C_l}{\bE C_{h,l}}\leq\frac{2\lambda+S}{S}$ and $\frac{2\bE C_h}{\bE C_{h,h}}\leq\frac{2\lambda+S}{S}$), then the WSPs will offer shared data plan for any $\alpha\in(0,1)$.
\end{itemize}
\end{pro}

The Proof of Proposition \ref{pro_sharealways} is given in Appendix D of the supplemental material.


\subsection{WSPs' Profits at Equilibrium Timing}

By substituting the equilibrium timing in Theorem \ref{thm_NEshare} to (\ref{equ_R_se}) and (\ref{equ_R_sl}), we can derive the equilibrium profits of the two WSPs. At the equilibrium share time, the profit of WSP $i, i=1,2$ is given as follows:
            \begin{itemize}
              \item The WSP $i$'s profit for offering shared data plan at the same time, i.e., $T_i^{ne}=T_j^{ne}=0$, is
              \end{itemize}
              \bee\label{equ_profitequalshare} \bar{R}_i^s=\eta_iN\cE\frac{1}{S}+\eta_i\eta_j N(\cD-\cE)\frac{1}{\lambda+S}+\frac{1}{2}N_0\cE(\frac{1}{S}-\frac{1}{\lambda_0+S}). \ene
              \begin{itemize}
              \item The WSP $i$'s profit for offering shared data plan first, i.e., $T_i^{ne}=0$ is
\bee\label{equ_shareprofitfirst}\begin{split} \bar{R}_i^{s,\leq}=&\cE N(\frac{1}{S}-\frac{1}{\lambda+S})+\eta_i^2N\frac{1}{\lambda+S}\cE\\
&+\eta_i\eta_jN\cD\frac{1}{\lambda+S}+N_0\cE(\frac{1}{S}-\frac{1}{\lambda_0+S})\\
&-(1-\eta_i)N\cE(\frac{1}{S}-\frac{1}{\lambda+S})(\frac{1}{\kappa_j^s})^{\frac{\lambda+S}{\lambda-\lambda_0}}\\
&-\frac{1}{2}N_0\cE(\frac{1}{S}-\frac{1}{\lambda_0+S})(\frac{1}{\kappa_j^s})^{\frac{\lambda_0+S}{\lambda-\lambda_0}}.
\end{split}\ene
              \item The WSP $i$'s profit for offering shared data plan late, i.e., $T_i^{ne}=\frac{\log\kappa_i^s}{\lambda-\lambda_0}$, is
\bee\label{equ_shareprofitlate}\begin{split} \bar{R}_i^{s,>}=&\eta_iN\cD\frac{1}{\lambda+S}(1-(\frac{1}{\kappa_i^s})^{\frac{\lambda+S}{\lambda-\lambda_0}})\\
&+\eta_iN\cE\frac{1}{S}(\frac{1}{\kappa_i^s})^{\frac{\lambda+S}{\lambda-\lambda_0}}\\
&+\eta_i\eta_j N(\cD-\cE)\frac{1}{\lambda+S}(\frac{1}{\kappa_i^s})^{\frac{\lambda+S}{\lambda-\lambda_0}}\\
&+\frac{1}{2}N_0\cE(\frac{1}{S}-\frac{1}{\lambda_0+S})(\frac{1}{\kappa_i^s})^{\frac{\lambda_0+S}{\lambda-\lambda_0}}.
\end{split}\ene
            \end{itemize}

Fig. \ref{NEprofitshare} numerically shows how the WSPs' equilibrium profits change with their market shares in the three $\eta_0$ regimes. As shown in Fig. \ref{largeprofitshare} and \ref{mediumprofitshare}, if both WSPs choose shared data immediately, each WSP's profit increases with its market share. However, as shown in Fig. \ref{mediumprofitshare}, a WSP's profit for earlier share may decrease with its market share. This is because WSP $j$ brings forward its upgrade time due to the reduced market share $\eta_j$, and thus less new users and WSP $j$'s users subscribe to WSP $i$. Similarly, in Fig. \ref{smallprofitshare}, the WSP $i$'s profit suddenly reduces as $\eta_i$ increases across the threshold point $1-\hat{\eta}_s$ (i.e., from medium to large market share). The reason is that WSP $i$ with large market share faces large revenue loss of overage charge, which cannot be made up by the revenue gain from the few new users.


\section{Conclusion}\label{sec_conclude}

This paper analytically studies the competitive WSPs' timing of offering innovative data plans. For the WSP with small market share,
it is better to upgrade first to attract more users (especially from the other WSP). While for the WSP with large market share, it is better to upgrade late to avoid
immediate loss in overage charge. For the WSPs with similar market shares, they should upgrade immediately for large new user population and may
not upgrade for small new user population. The launching of different innovative data plans also depends on their strategies. According to the composition of users in a family, the WSPs may upgrade to shared data plan simultaneously even when no new user arrives, which is different from the rollover data plan.


In the future, we plan to extend our results in two possible directions. First, we could consider a variety of separate rollover and shared data plans (e.g., for 2G/3G/4G) to fit heterogeneous users. Though the analysis is more involved by modeling mixed user churn flows from any existing plan to any innovative plan under the same or different WSPs, we still expect the WSP of small market share chooses innovative data plan first. Second, we can extend users' static data usage model and consider that the innovative data plans may motivate users to increase their usage. In this case, the WSPs' overage charges reduce less severely and they have more incentives to announce innovative plans earlier.



\appendices
\section{Proof of Theorem~3.1}

According to Proposition 3.1, we can see that the possible equilibrium rollover time for the WSPs are ($T_i^{ne}=T_j^{ne}=0$), ($T_i^{ne}=T_j^{ne}>0$), ($T_i^{ne}=0, T_j^{ne}=\frac{\log\kappa_j}{\lambda-\lambda_0}$), ($T_j^{ne}=0, T_i^{ne}=\frac{\log\kappa_i}{\lambda-\lambda_0}$), ($T_i^{ne}=T_j^{ne}=\infty$).
In the following, we will discuss the conditions for each equilibrium.

($T_i^{ne}=T_j^{ne}=0$) is the equilibrium if and only if
\bee\label{equ_con1}\begin{split} \frac{\partial R_i^{r,>}(T_i,0)}{\partial T_i}=&2\alpha N_i(\bE C_h-\bE C_h^r\frac{\lambda+S}{S})e^{\lambda T_j-(\lambda+S)T_i}\\
&-\alpha N_0\bE C_h^r\frac{\lambda_0}{S}e^{\lambda_0 T_j-(\lambda_0+S)T_i}\leq 0, \end{split}\ene
and \bee\label{equ_con2}\begin{split} \frac{\partial R_j^{r,>}(0, T_j)}{\partial T_j}=&2\alpha N_j(\bE C_h-\bE C_h^r\frac{\lambda+S}{S})e^{\lambda T_i-(\lambda+S)T_j}\\
&-\alpha N_0\bE C_h^r\frac{\lambda_0}{S}e^{\lambda_0 T_i-(\lambda_0+S)T_j}\leq 0, \end{split}\ene
i.e.,
$\kappa_i\leq 1$ and $\kappa_j\leq 1$.

It is manifest that when $\bE C_h\leq\bE C_h^r\frac{\lambda+S}{S}$, which means the heavy user's expected cost after rollover is close to that before rollover, both WSPs will rollover at the beginning.

Then, we show that if $T_i^{ne}=T_j^{ne}$ is the equilibrium, we have $T_i^{ne}=T_j^{ne}=0$.

If $T_i^{ne}=T_j^{ne}$ is the equilibrium, if and only if $\kappa_i\leq 1$, $\kappa_j\leq 1$ and
\bee \frac{\partial R_i^{r,\leq}}{\partial T_i}|_{T_i=T_j}\geq 0, \ene
and \bee \frac{\partial R_j^{r,\leq}}{\partial T_j}|_{T_j=T_i}\geq 0, \ene
i.e.,
\bee\label{equ_con3}\begin{split} &2\alpha(N_i\bE C_h-N\bE C_h^r+N_j\frac{S}{\lambda+S}\bE C_h^r)-2\alpha N_0\frac{\lambda_0}{\lambda_0+S}\bE C_h^r\\&-2\alpha N_j(\frac{\lambda}{S}-\frac{\lambda}{\lambda+S})\bE C_h^r+\alpha N_0(\frac{\lambda_0}{\lambda_0+S}-\frac{\lambda_0}{S})\bE C_h^r\geq 0, \end{split}\ene
and
\bee\label{equ_con4}\begin{split} &2\alpha(N_j\bE C_h-N\bE C_h^r+N_i\frac{S}{\lambda+S}\bE C_h^r)-2\alpha N_0\frac{\lambda_0}{\lambda_0+S}\bE C_h^r\\&-2\alpha N_i(\frac{\lambda}{S}-\frac{\lambda}{\lambda+S})\bE C_h^r+\alpha N_0(\frac{\lambda_0}{\lambda_0+S}-\frac{\lambda_0}{S})\bE C_h^r\geq 0, \end{split}\ene

According to $\kappa_i\leq 1$ and $\kappa_j\leq 1$, the proportion of new users should satisfy $\frac{S}{\lambda_0}(\frac{\bE C_h}{\bE C_h^r}-\frac{\lambda+S}{S})\leq\eta_0<\frac{2S}{\lambda_0}(\frac{\bE C_h}{\bE C_h^r}-\frac{\lambda+S}{S})$. Under this circumstance, by solving (\ref{equ_con3}) and (\ref{equ_con4}), we have $\eta_i>0.5$ and $\eta_j=1-\eta_i>0.5$, which can't be satisfied simultaneously. Therefore, $T_i^{ne}=T_j^{ne}=0$.

The necessary condition for ($T_i^{ne}=0, T_j^{ne}=\frac{\log\kappa_j}{\lambda-\lambda_0}$) is $\kappa_j>1$ and
$\frac{\partial R_i^{r,\leq}}{\partial T_i}|_{T_j=\frac{\log\kappa_j}{\lambda-\lambda_0}}<0$, i.e., $g(\eta_i)<0$, where
\bee\begin{split} g(\eta_i)=&2\alpha(N_i\bE C_h-N\bE C_h^r+N_j\frac{S}{\lambda+S}\bE C_h^r)\\
&-2\alpha N_j(\frac{\lambda}{S}-\frac{\lambda}{\lambda+S})\bE C_h^r(\frac{1}{\kappa_j})^{\frac{\lambda+S}{\lambda-\lambda_0}}\\
&-\alpha N_0(\frac{\lambda_0}{S}-\frac{\lambda_0}{\lambda_0+S})\bE C_h^r(\frac{1}{\kappa_j})^{\frac{\lambda_0+S}{\lambda-\lambda_0}}\\
&-2\alpha N_0\frac{\lambda_0}{\lambda_0+S}\bE C_h^r. \end{split}\ene
And the necessary condition for ($T_j^{ne}=0, T_i^{ne}=\frac{\log\kappa_i}{\lambda-\lambda_0}$) is $\kappa_i>1$ and $\frac{\partial R_j^{r,e}}{\partial T_j}|_{T_i=\frac{\log\kappa_i}{\lambda-\lambda_0}}<0$, i.e.,
\bee\begin{split} &2\alpha(N_j\bE C_h-N\bE C_h^r+N_i\frac{S}{\lambda+S}\bE C_h^r)-2\alpha N_0\frac{\lambda_0}{\lambda_0+S}\bE C_h^r\\
&-2\alpha N_i(\frac{\lambda}{S}-\frac{\lambda}{\lambda+S})\bE C_h^r(\frac{1}{\kappa_i})^{\frac{\lambda+S}{\lambda-\lambda_0}}\\
&+\alpha N_0(\frac{\lambda_0}{\lambda_0+S}-\frac{\lambda_0}{S})\bE C_h^r(\frac{1}{\kappa_i})^{\frac{\lambda_0+S}{\lambda-\lambda_0}}<0. \end{split}\ene

Otherwise, both WSPs will not rollover, i.e., ($T_i^{ne}=T_j^{ne}=\infty$).


When $\bE C_h>\bE C_h^r\frac{\lambda+S}{S}$, according to $\kappa_i\leq 1$ and $\kappa_j\leq 1$, $T_i^{ne}=T_j^{ne}=0$ is the equilibrium if and only if $1-\frac{N_0\bE C_h^r\frac{\lambda_0}{S}}{2N(\bE C_h-\bE C_h^r\frac{\lambda+S}{S})}\leq\eta_i\leq\frac{N_0\bE C_h^r\frac{\lambda_0}{S}}{2N(\bE C_h-\bE C_h^r\frac{\lambda+S}{S})}$. Therefore, when $\eta_0\geq \frac{2S}{\lambda_0}(\frac{\bE C_h}{\bE C_h^r}-\frac{\lambda+S}{S})$, we have $\frac{N_0\bE C_h^r\frac{\lambda_0}{S}}{2N(\bE C_h-\bE C_h^r\frac{\lambda+S}{S})}\geq 1$, which means the condition always holds and $T_i^{ne}=T_j^{ne}=0$ is the equilibrium.

Since $g(\eta_i=0)<0$ and $g(\eta_i)$ decreases with $\eta_0$, the solution of $g(\eta_i=0)=0$, denoted as $\bar{\eta}_r$, increases with $\eta_0$. Therefore, if $g(\eta_i=0.5)<0$ for a given $\hat{\eta}_0$, then, for any $\eta_0>\hat{\eta}_0$, $g(\eta_i=0.5)<0$, which means $\bar{\eta}_r>0.5$. When $\eta_0=\frac{S}{\lambda_0}(\frac{\bE C_h}{\bE C_h^r}-\frac{\lambda+S}{S})$ and $\eta_i=0.5$, we have $\kappa_j=1$. Since $\bE C_h>\bE C_h^r\frac{\lambda+S}{S}$, it is easy to check that $g(\eta_i=0.5)<0$. Therefore, when $\frac{S}{\lambda_0}(\frac{\bE C_h}{\bE C_h^r}-\frac{\lambda+S}{S})<\eta_0<\frac{2S}{\lambda_0}(\frac{\bE C_h}{\bE C_h^r}-\frac{\lambda+S}{S})$, i.e., $0.5<\frac{N_0\bE C_h^r\frac{\lambda_0}{S}}{2N(\bE C_h-\bE C_h^r\frac{\lambda+S}{S})}<1$, $\bar{\eta}_r>0.5>1-\frac{N_0\bE C_h^r\frac{\lambda_0}{S}}{2N(\bE C_h-\bE C_h^r\frac{\lambda+S}{S})}$. Then, when $0\leq\eta_i<1-\frac{N_0\bE C_h^r\frac{\lambda_0}{S}}{2N(\bE C_h-\bE C_h^r\frac{\lambda+S}{S})}$, $g(\eta_i)\leq 0$ is always satisfied and thus $T_i^{ne}=0, T_j^{ne}=\frac{\log\kappa_j}{\lambda-\lambda_0}$ is the equilibrium. Similarly, for the case when $\frac{N_0\bE C_h^r\frac{\lambda_0}{S}}{2N(\bE C_h-\bE C_h^r\frac{\lambda+S}{S})}<\eta_i\leq 1$, we have $T_j^{ne}=0, T_i^{ne}=\frac{\log\kappa_i}{\lambda-\lambda_0}$ is the equilibrium. Note that when $0.5<\frac{N_0\bE C_h^r\frac{\lambda_0}{S}}{2N(\bE C_h-\bE C_h^r\frac{\lambda+S}{S})}<1$, $1-\frac{N_0\bE C_h^r\frac{\lambda_0}{S}}{2N(\bE C_h-\bE C_h^r\frac{\lambda+S}{S})}<\frac{N_0\bE C_h^r\frac{\lambda_0}{S}}{2N(\bE C_h-\bE C_h^r\frac{\lambda+S}{S})}$. Thus, when $1-\frac{N_0\bE C_h^r\frac{\lambda_0}{S}}{2N(\bE C_h-\bE C_h^r\frac{\lambda+S}{S})}\leq\eta_i\leq\frac{N_0\bE C_h^r\frac{\lambda_0}{S}}{2N(\bE C_h-\bE C_h^r\frac{\lambda+S}{S})}$, $T_i^{ne}=T_j^{ne}=0$.

Define $\phi(\eta_0):=g(\eta_i=0.5)$ and denote its solution as $\bar{\eta}_0$. Since $\bE C_h>\bE C_h^r\frac{\lambda+S}{S}$, it is easy to check that $\phi(\eta_0=0)>0$, and thus $\bar{\eta}_0>0$. Then, when $\eta_0\leq\bar{\eta}_0$, we have $\bar{\eta}_r\leq 0.5$ and when $\eta_0>\bar{\eta}_0$, we have $\bar{\eta}_r>0.5$. We can check that $\phi(\eta_0=\frac{S}{\lambda_0}(\frac{\bE C_h}{\bE C_h^r}-\frac{\lambda+S}{S}))<0$ when $\bE C_h>\bE C_h^r\frac{\lambda+S}{S}$, which results in $\bar{\eta}_0<\frac{S}{\lambda_0}(\frac{\bE C_h}{\bE C_h^r}-\frac{\lambda+S}{S})$. When $\bar{\eta}_0<\eta_0\leq\frac{S}{\lambda_0}(\frac{\bE C_h}{\bE C_h^r}-\frac{\lambda+S}{S})$, $\frac{N_0\bE C_h^r\frac{\lambda_0}{S}}{2N(\bE C_h-\bE C_h^r\frac{\lambda+S}{S})}\leq 1-\frac{N_0\bE C_h^r\frac{\lambda_0}{S}}{2N(\bE C_h-\bE C_h^r\frac{\lambda+S}{S})}$. Thus, when $\max(1-\bar{\eta}_r, \frac{N_0\bE C_h^r\frac{\lambda_0}{S}}{2N(\bE C_h-\bE C_h^r\frac{\lambda+S}{S})})<\eta_i<\min(\bar{\eta}_r, 1-\frac{N_0\bE C_h^r\frac{\lambda_0}{S}}{2N(\bE C_h-\bE C_h^r\frac{\lambda+S}{S})})$, all the conditions $\kappa_i>1$, $\kappa_j>1$, $\frac{\partial R_j^{r,\leq}}{\partial T_j}|_{T_i=\frac{\log\kappa_i}{\lambda-\lambda_0}}<0$ and $\frac{\partial R_i^{r,\leq}}{\partial T_i}|_{T_j=\frac{\log\kappa_j}{\lambda-\lambda_0}}<0$ are satisfied and the WSPs choose to rollover early or late by comparing their profits. For any $\eta_0\in(\bar{\eta}_0, \frac{S}{\lambda_0}(\frac{\bE C_h}{\bE C_h^r}-\frac{\lambda+S}{S}))$, denote $\tilde{\eta}_r$ as the solution of $\bar{R}_i^{r,\leq}(\eta_i)-\bar{R}_i^{r,>}(\eta_i)=0$. Then, when WSP $i$'s market share is large ($\eta_i>\tilde{\eta}_r$), it chooses to rollover late ($\bar{R}_i^{r,\leq}(\eta_i)<\bar{R}_i^{r,>}(\eta_i)$); when WSP $i$'s market share is small ($\eta_i<\tilde{\eta}_r$), it chooses to rollover early ($\bar{R}_i^{r,\leq}(\eta_i)>\bar{R}_i^{r,>}(\eta_i)$). Therefore, when $\eta_i<\tilde{\eta}_r$ and $\eta_j>\tilde{\eta}_r$, i.e., $\eta_i<\min(\tilde{\eta}_r,1-\tilde{\eta}_r)$, $T_i^{ne}=0, T_j^{ne}=\frac{\log\kappa_j}{\lambda-\lambda_0}$ is the equilibrium and when $\eta_j<\tilde{\eta}_r$ and $\eta_i>\tilde{\eta}_r$, i.e., $\eta_i>\max(\tilde{\eta}_r, 1-\tilde{\eta}_r)$, $T_i^{ne}=0, T_j^{ne}=\frac{\log\kappa_j}{\lambda-\lambda_0}$ is the equilibrium.
If $\tilde{\eta}_r\geq 0.5$, the conditions $\eta_i<\tilde{\eta}_r, \eta_j<\tilde{\eta}_r$ are satisfied when $1-\tilde{\eta}_r<\eta_i<\tilde{\eta}_r$, then, both WSPs choose to rollover early ($T_i^{ne}=T_j^{ne}=0$); If $\tilde{\eta}_r<0.5$, the conditions $\eta_i>\tilde{\eta}_r, \eta_j>\tilde{\eta}_r$ are satisfied when $\tilde{\eta}_r<\eta_i<1-\tilde{\eta}_r$, then, both WSPs choose to rollover late ($T_i^{ne}=T_j^{ne}=\infty$). According to WSP $i$'s profits for first and late rollover, i.e.,
\bee\label{equ_profitearly}\begin{split} \bar{R}_i^{r,\leq}=&2\alpha N\bE C_h^r\frac{1}{S}-2\alpha N_j\bE C_h^r\frac{1}{S+\lambda}\\
&-2\alpha N_j\bE C_h^r(\frac{1}{S}-\frac{1}{\lambda+S})(\frac{1}{\kappa_j})^{\frac{\lambda+S}{\lambda-\lambda_0}}\\
&+2\alpha N_0(\frac{1}{S}-\frac{1}{S+\lambda_0})\bE C_h^r\\
&-\alpha N_0(\frac{1}{S}-\frac{1}{S+\lambda_0})(\frac{1}{\kappa_j})^{\frac{\lambda_0+S}{\lambda-\lambda_0}}\bE C_h^r, \end{split}\ene
and
\bee\label{equ_profitlate}\begin{split} \bar{R}_i^{r,>}=&2\alpha N_i\frac{1}{\lambda+S}\bE C_h(1-(\frac{1}{\kappa_i})^{\frac{\lambda+S}{\lambda-\lambda_0}})\\
&+2\alpha N_i\frac{1}{S}\bE C_h^r(\frac{1}{\kappa_i})^{\frac{\lambda+S}{\lambda-\lambda_0}}\\
&+\alpha N_0(\frac{1}{S}-\frac{1}{S+\lambda_0})(\frac{1}{\kappa_i})^{\frac{\lambda_0+S}{\lambda-\lambda_0}}\bE C_h^r, \end{split}\ene
when $\eta_0\in(\bar{\eta}_0, \frac{S}{\lambda_0}(\frac{\bE C_h}{\bE C_h^r}-\frac{\lambda+S}{S}))$, $\bar{R}_i^{r,\leq}(\eta_i)-\bar{R}_i^{r,>}(\eta_i)$ decreases with $\eta_i$. Define $\psi(\eta_0):=\bar{R}_i^{r,\leq}(\eta_i=0.5)-\bar{R}_i^{r,>}(\eta_i=0.5)$. When $\bE C_h<\bE C_h^r\frac{2\lambda+S}{S}$, we can check that $\psi(\eta_0)\geq 0$ always holds for $\eta_0\in(0, \frac{S}{\lambda_0}(\frac{\bE C_h}{\bE C_h^r}-\frac{\lambda+S}{S}))$, which means $\tilde{\eta}_r\geq 0.5$ for $\eta_0\in(\bar{\eta}_0, \frac{S}{\lambda_0}(\frac{\bE C_h}{\bE C_h^r}-\frac{\lambda+S}{S}))$.


Note that $T_i^{ne}=0, T_j^{ne}=\frac{\log\kappa_j}{\lambda-\lambda_0}$ is the equilibrium if $\eta_i<\min(\bar{\eta}_r, 1-\frac{N_0\bE C_h^r\frac{\lambda_0}{S}}{2N(\bE C_h-\bE C_h^r\frac{\lambda+S}{S})})$ and $T_j^{ne}=0, T_i^{ne}=\frac{\log\kappa_i}{\lambda-\lambda_0}$ is the equilibrium if $\eta_i>\max(1-\bar{\eta}_r, \frac{N_0\bE C_h^r\frac{\lambda_0}{S}}{2N(\bE C_h-\bE C_h^r\frac{\lambda+S}{S})})$. For $\eta_0\leq\bar{\eta}_0$, as $\bar{\eta}_r\leq 0.5$ and $\frac{N_0\bE C_h^r\frac{\lambda_0}{S}}{2N(\bE C_h-\bE C_h^r\frac{\lambda+S}{S})}<0.5$, then, we have $T_i^{ne}=0, T_j^{ne}=\frac{\log\kappa_j}{\lambda-\lambda_0}$ when $0\leq\eta_i\leq\bar{\eta}_r$ and $T_j^{ne}=0, T_i^{ne}=\frac{\log\kappa_i}{\lambda-\lambda_0}$ when $1-\bar{\eta}_r\leq\eta_i\leq 1$. When $\bar{\eta}_r<\eta_i<1-\bar{\eta}_r$, all the equilibrium conditions can't be satisfied and thus $T_i^{ne}=T_j^{ne}=\infty$. \qed

\section{Proof of Proposition~3.2}\label{app_pro_rollover_threshold}

When $\bE C_h\leq\bE C_h^r\frac{\lambda+S}{S}$, the WSPs rollover at the same time. If $\bar{R}_i^r(\eta_i=1)\geq R_i(\eta_i=1)$, which means the cost reduction $\bE C_h-\bE C_h^r$ is very small and the number of new users $N_0$ is large, the WSPs always gain profit by rollover data. Otherwise, there exists an unique threshold $\eta_i^{th}$ such that when $\eta_i\leq \eta_i^{th}$, WSP $i$ gains profit and when $\eta_i>\eta_i^{th}$, WSP $i$ losses profit by rollover data.

In the following, we discuss the situation when $\bE C_h>\bE C_h^r\frac{\lambda+S}{S}$.
When $\eta_i=0$, the WSP's profit before rollover is $0$, which is smaller than its profit after rollover.

As $\bar{R}_i^r$ increases with $\eta_0$, we have
\bee\begin{split} \bar{R}_i^r(\eta_i=1)\leq&2\alpha N\bE C_h^r\frac{1}{S}+\alpha N(\frac{1}{S}-\frac{1}{S+\lambda_0})\bE C_h^r\\
<&2\alpha N\bE C_h^r\frac{\lambda+S}{S}\frac{1}{S}\leq2\alpha N\bE C_h\frac{1}{S}\\
=&R_i(\eta_i=1). \end{split}\ene
Thus, when $\eta_i=1$, the WSP's profit for rollover at the same time is smaller than its profit before rollover. 

By noting that the WSP's profit linearly increases with $\eta_i$, there is only one intersection of $\bar{R}_i^r(\eta_i)$ and $R_i(\eta_i)$ for large $\eta_0$.

For any $\eta_i$, by comparing $\bar{R}_i^r$ and $\bar{R}_i^{r,\leq}$, it is obvious that $\bar{R}_i^r<\bar{R}_i^{r,\leq}$.


Note that
\bee\begin{split} \bar{R}_i^{r,>}(\eta_i=1)<&2\alpha N\frac{1}{\lambda+S}\bE C_h+\alpha N(\frac{1}{S}-\frac{1}{S+\lambda_0})\bE C_h^r\\
 <&2\alpha N\bE C_h\frac{1}{S}=R_i(\eta_i=1). \end{split}\ene

For medium $\eta_0$, if the WSP $i$ chooses to rollover late, then $\eta_i>0.5$. Therefore, we have $\bar{R}_i^{r,>}(\eta_i)<R_i$ due to $\eta_0<1$ and $\bE C_h>\bE C_h^r\frac{\lambda+S}{S}$.

Note that $R_i(\eta_i)$ increases faster than $\bar{R}_i^r(\eta_i)$. Therefore, there is an unique intersection of the WSP's profits before and after rollover for medium $\eta_0$.

For small $\eta_0$, when $\bar{\eta}_r<\eta_i<1-\bar{\eta}_r$, the WSPs choose not to rollover data. Thus, the unique intersection of the WSP's profits before and after rollover is the $\eta_i^{th}$ such that $R_i(\eta_i)=\bar{R}_i^{r,\leq}(\eta_i)$. \qed

\section{Proof of Theorem~4.1}

\begin{figure}[ht]
\centering\includegraphics[scale=0.4]{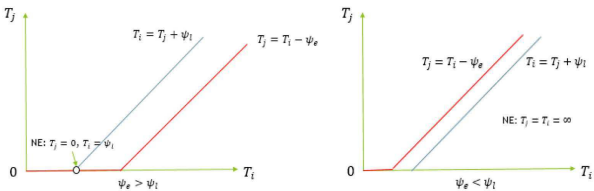}\caption{Two WSPs' best responses to each other and the equilibrium rollover time.}\label{prove}
\end{figure}

Note that if a WSP offers shared data plan first and the other offers later, the equilibrium will converge to ($T_i^{ne}=0, T_j^{ne}>0$) or ($T_i^{ne}=T_j^{ne}=\infty$) as shown in Fig. \ref{prove}. Therefore, the possible equilibrium share time for the WSPs are ($T_i^{ne}=T_j^{ne}=0$), ($T_i^{ne}=0, T_j^{ne}=\frac{\log\kappa_j^s}{\lambda-\lambda_0}$), ($T_j^{ne}=0, T_i^{ne}=\frac{\log\kappa_i^s}{\lambda-\lambda_0}$), and ($T_i^{ne}=T_j^{ne}=\infty$), where $T_i^{ne}=\frac{\log\kappa_i^s}{\lambda-\lambda_0}$ is obtained by letting $\frac{\partial R_i^{s,>}}{\partial T_i}|_{T_j=0}=0$.

($T_i^{ne}=T_j^{ne}=0$) is the equilibrium if and only if $\frac{\partial R_i^{s,>}(T_i,0)}{\partial T_i}\leq 0$ and $\frac{\partial R_j^{s,>}(T_j,0)}{\partial T_j}\leq 0$, i.e., both $\eta_i, \eta_j$ satisfies
\bee\label{equ_shareprove1} \eta_i, \eta_j\leq \frac{\cE\frac{\lambda}{S}+\sqrt{(\cE\frac{\lambda}{S})^2+2(\cD-\cE)\cE\eta_0\frac{\lambda_0}{S}}}{2(\cD-\cE)}. \ene

Note the if $\cD\leq\cE\frac{\lambda+S}{S}$ or $\eta_0\geq \frac{2(\cD S-\cE(\lambda+S))}{\cE\lambda_0}$, (\ref{equ_shareprove1}) is always satisfied due to $\frac{\cE\frac{\lambda}{S}+\sqrt{(\cE\frac{\lambda}{S})^2+2(\cD-\cE)\cE\eta_0\frac{\lambda_0}{S}}}{2(\cD-\cE)}\geq 1$.

In the following, we discuss the situation when $\cD>\cE\frac{\lambda+S}{S}$ and $\eta_0<\frac{2(\cD S-\cE(\lambda+S))}{\cE\lambda_0}$.

($T_i^{ne}=0, T_j^{ne}=\frac{\log\kappa_j^s}{\lambda-\lambda_0}$) is the equilibrium if $\kappa_j^s>1$ and $\frac{\partial R_i^{s,\leq}}{\partial T_i}|_{T_j=\frac{\log\kappa_j^s}{\lambda-\lambda_0}}<0$, i.e.,
\bee\begin{split} v(\eta_i)=&\cD\eta_i(\frac{\lambda}{\lambda+S}+\eta_i\frac{S}{\lambda+S})-\cE(\frac{\lambda}{\lambda+S}+\eta_i^2\frac{S}{\lambda+S})\\
&-\cE\eta_0\frac{\lambda_0}{\lambda_0+S}-\eta_j\cE(\frac{\lambda}{S}-\frac{\lambda}{\lambda+S})(\frac{1}{\kappa_j^s})^{\frac{\lambda+S}{\lambda-\lambda_0}}\\
&-\frac{1}{2}\cE\eta_0\frac{\lambda_0^2}{S(\lambda_0+S)}(\frac{1}{\kappa_j^s})^{\frac{\lambda_0+S}{\lambda-\lambda_0}}<0. \end{split}\ene

Denote $\hat{\eta}_s$ as the solution to $v(\eta_i)=0$ and $\bar{\eta}_0^s$ as the solution to $\chi(\eta_0):=v(\eta_i=0.5)=0$. Note that $v(\eta_i=0)<0$ and $v(\eta_i)$ decreases with $\eta_0$, thus $\hat{\eta}_s$ increases with $\eta_0$. Then, when $\eta_0\leq \bar{\eta}_0^s$, $\hat{\eta}_s\leq 0.5$ and when $\eta_0>\bar{\eta}_0^s$, $\hat{\eta}_s>0.5$. Therefore, if $0\leq\eta_0\leq\min((\bar{\eta}_0^s)^+,(\frac{\cD S-\cE(2\lambda+S)}{2\cE\lambda_0})^+)$, then, when $0\leq\eta_i\leq\hat{\eta}_s$, $T_i^{ne}=0, T_j^{ne}=\frac{\log\kappa_j^s}{\lambda-\lambda_0}$; When $\hat{\eta}_s<\eta_i<1-\hat{\eta}_s$, $T_i^{ne}=T_j^{ne}=\infty$; When $1-\hat{\eta}_s\leq\eta_i\leq 1$, $T_j^{ne}=0, T_i^{ne}=\frac{\log\kappa_i^s}{\lambda-\lambda_0}$.

Then, we consider the case when $\min((\bar{\eta}_0^s)^+,(\frac{\cD S-\cE(2\lambda+S)}{2\cE\lambda_0})^+)<\eta_0<\frac{2(\cD S-\cE(\lambda+S))}{\cE\lambda_0}$. If $\min(\hat{\eta}_s, 1-\frac{\cE\frac{\lambda}{S}+\sqrt{(\cE\frac{\lambda}{S})^2+2(\cD-\cE)\cE\eta_0\frac{\lambda_0}{S}}}{2(\cD-\cE)})\leq\max(1-\hat{\eta}_s, \frac{\cE\frac{\lambda}{S}+\sqrt{(\cE\frac{\lambda}{S})^2+2(\cD-\cE)\cE\eta_0\frac{\lambda_0}{S}}}{2(\cD-\cE)})$, according to the above analysis, we can conclude that when $0\leq\eta_i<\min(\hat{\eta}_s, 1-\frac{\cE\frac{\lambda}{S}+\sqrt{(\cE\frac{\lambda}{S})^2+2(\cD-\cE)\cE\eta_0\frac{\lambda_0}{S}}}{2(\cD-\cE)})$, $T_i^{ne}=0, T_j^{ne}=\frac{\log\kappa_j^s}{\lambda-\lambda_0}$; when $\min(\hat{\eta}_s, 1-\frac{\cE\frac{\lambda}{S}+\sqrt{(\cE\frac{\lambda}{S})^2+2(\cD-\cE)\cE\eta_0\frac{\lambda_0}{S}}}{2(\cD-\cE)})\leq\eta_i\leq\max(1-\hat{\eta}_s, \frac{\cE\frac{\lambda}{S}+\sqrt{(\cE\frac{\lambda}{S})^2+2(\cD-\cE)\cE\eta_0\frac{\lambda_0}{S}}}{2(\cD-\cE)})$, $T_i^{ne}=T_j^{ne}=0$; when $\max(1-\hat{\eta}_s, \frac{\cE\frac{\lambda}{S}+\sqrt{(\cE\frac{\lambda}{S})^2+2(\cD-\cE)\cE\eta_0\frac{\lambda_0}{S}}}{2(\cD-\cE)})<\eta_i\leq 1$, $T_j^{ne}=0, T_i^{ne}=\frac{\log\kappa_i^s}{\lambda-\lambda_0}$. If $\min(\hat{\eta}_s, 1-\frac{\cE\frac{\lambda}{S}+\sqrt{(\cE\frac{\lambda}{S})^2+2(\cD-\cE)\cE\eta_0\frac{\lambda_0}{S}}}{2(\cD-\cE)})>\max(1-\hat{\eta}_s, \frac{\cE\frac{\lambda}{S}+\sqrt{(\cE\frac{\lambda}{S})^2+2(\cD-\cE)\cE\eta_0\frac{\lambda_0}{S}}}{2(\cD-\cE)})$, all the conditions \bee \eta_i, \eta_j>\frac{\cE\frac{\lambda}{S}+\sqrt{(\cE\frac{\lambda}{S})^2+2(\cD-\cE)\cE\eta_0\frac{\lambda_0}{S}}}{2(\cD-\cE)}, \ene $\frac{\partial R_j^{s,\leq}}{\partial T_j}|_{T_i=\frac{\log\kappa_i^s}{\lambda-\lambda_0}}<0$ and $\frac{\partial R_i^{s,\leq}}{\partial T_i}|_{T_j=\frac{\log\kappa_j^s}{\lambda-\lambda_0}}<0$ are satisfied and the WSPs choose to offer shared data plan early or late by comparing their profits. Denote $\tilde{\eta}_s$ as the solution of $\bar{R}_i^{s,\leq}(\eta_i)-\bar{R}_i^{s,>}(\eta_i)=0$. WSP $i$ choose shared data plan earlier if $\eta_i\leq\tilde{\eta}_s$ and later if $\eta_i>\tilde{\eta}_s$. Since $\cD\geq\cE\frac{4\lambda+S}{S}$, we have $\tilde{\eta}_s\geq 0.5$. Thus, when $\eta_i<1-\tilde{\eta}_s$, WSP $i$ prefers to offer shared data plan first while WSP $j$ prefers later, and when $\eta_i>\tilde{\eta}_s$, WSP $j$ prefers earlier while WSP $i$ prefers later. When $1-\tilde{\eta}_s\leq\eta_i\leq\tilde{\eta}_s$, both WSPs prefer to offer shared data plan earlier and thus $T_i^{ne}=T_j^{ne}=0$. Therefore, for the case $\min((\bar{\eta}_0^s)^+,(\frac{\cD S-\cE(2\lambda+S)}{2\cE\lambda_0})^+)<\eta_0<\frac{2(\cD S-\cE(\lambda+S))}{\cE\lambda_0}$, we can conclude that when $0\leq\eta_i<\underline{\eta_s}$, $T_i^{ne}=0, T_j^{ne}=\frac{\log\kappa_j^s}{\lambda-\lambda_0}$; when $\underline{\eta_s}\leq\eta_i\leq 1-\underline{\eta_s}$, $T_i^{ne}=T_j^{ne}=0$; when $1-\underline{\eta_s}<\eta_i\leq 1$, $T_j^{ne}=0, T_i^{ne}=\frac{\log\kappa_i^s}{\lambda-\lambda_0}$. \qed

\section{Proof of Proposition~4.1}\label{app_pro_sharealways}

We only need to check the conditions when $\cD S\leq\cE(2\lambda+S)$, i.e.,
\bee\label{equ_prove2}\begin{split} &2S(\bE C_h+\bE C_l)-2(2\lambda+S)\bE C_{h,l}\\
\leq &\alpha(2S\bE C_{l}+(2\lambda+S)\bE C_{h,h}-2(2\lambda+S)\bE C_{h,l}). \end{split}\ene

If $\frac{\bE C_h+\bE C_l}{\bE C_{h,l}}>\frac{2\lambda+S}{S}$ and $\frac{2\bE C_h}{\bE C_{h,h}}\leq\frac{2\lambda+S}{S}$, we have \been\begin{split} &2S\bE C_{l}+(2\lambda+S)\bE C_{h,h}-2(2\lambda+S)\bE C_{h,l}\\
\geq &2S(\bE C_h+\bE C_l)-2(2\lambda+S)\bE C_{h,l}>0. \end{split}\enen
Thus, if $\alpha\geq\frac{2(\bE C_h+\bE C_l-\frac{2\lambda+S}{S}\bE C_{h,l})}{2\bE C_l+(\bE C_{h,h}-2\bE C_{h,l})\frac{2\lambda+S}{S}}$, (\ref{equ_prove2}) is satisfied.

If $\frac{\bE C_h+\bE C_l}{\bE C_{h,l}}\leq\frac{2\lambda+S}{S}$ and $\frac{2\bE C_h}{\bE C_{h,h}}>\frac{2\lambda+S}{S}$, we have \been\begin{split} &2S\bE C_{l}+(2\lambda+S)\bE C_{h,h}-2(2\lambda+S)\bE C_{h,l}\\
<&2S(\bE C_h+\bE C_l)-2(2\lambda+S)\bE C_{h,l}\leq 0. \end{split}\enen
Thus, if $\alpha\leq\frac{2(\bE C_h+\bE C_l-\frac{2\lambda+S}{S}\bE C_{h,l})}{2\bE C_l+(\bE C_{h,h}-2\bE C_{h,l})\frac{2\lambda+S}{S}}$, (\ref{equ_prove2}) is satisfied.

If $\frac{\bE C_h+\bE C_l}{\bE C_{h,l}}\leq\frac{2\lambda+S}{S}$ and $\frac{2\bE C_h}{\bE C_{h,h}}\leq\frac{2\lambda+S}{S}$, we have
$2S(\bE C_h+\bE C_l)-2(2\lambda+S)\bE C_{h,l}\leq 0$ and \been\begin{split} &2S\bE C_{l}+(2\lambda+S)\bE C_{h,h}-2(2\lambda+S)\bE C_{h,l}\\
\geq&2S(\bE C_h+\bE C_l)-2(2\lambda+S)\bE C_{h,l}. \end{split}\enen
Since $\alpha\in(0,1)$, (\ref{equ_prove2}) always holds.

If $\frac{\bE C_h+\bE C_l}{\bE C_{h,l}}>\frac{2\lambda+S}{S}$ and $\frac{2\bE C_h}{\bE C_{h,h}}>\frac{2\lambda+S}{S}$, we have we have
$2S(\bE C_h+\bE C_l)-2(2\lambda+S)\bE C_{h,l}>0$ and \been\begin{split} &2S\bE C_{l}+(2\lambda+S)\bE C_{h,h}-2(2\lambda+S)\bE C_{h,l}\\
<&2S(\bE C_h+\bE C_l)-2(2\lambda+S)\bE C_{h,l}. \end{split}\enen
By noting that $\alpha\in(0,1)$, (\ref{equ_prove2}) never holds. \qed

%

%
%
%




\bibliographystyle{IEEEtran}
\bibliography{infocom2018}

\begin{IEEEbiography}
[{\includegraphics[width=1in,height=1.25in,clip,keepaspectratio]{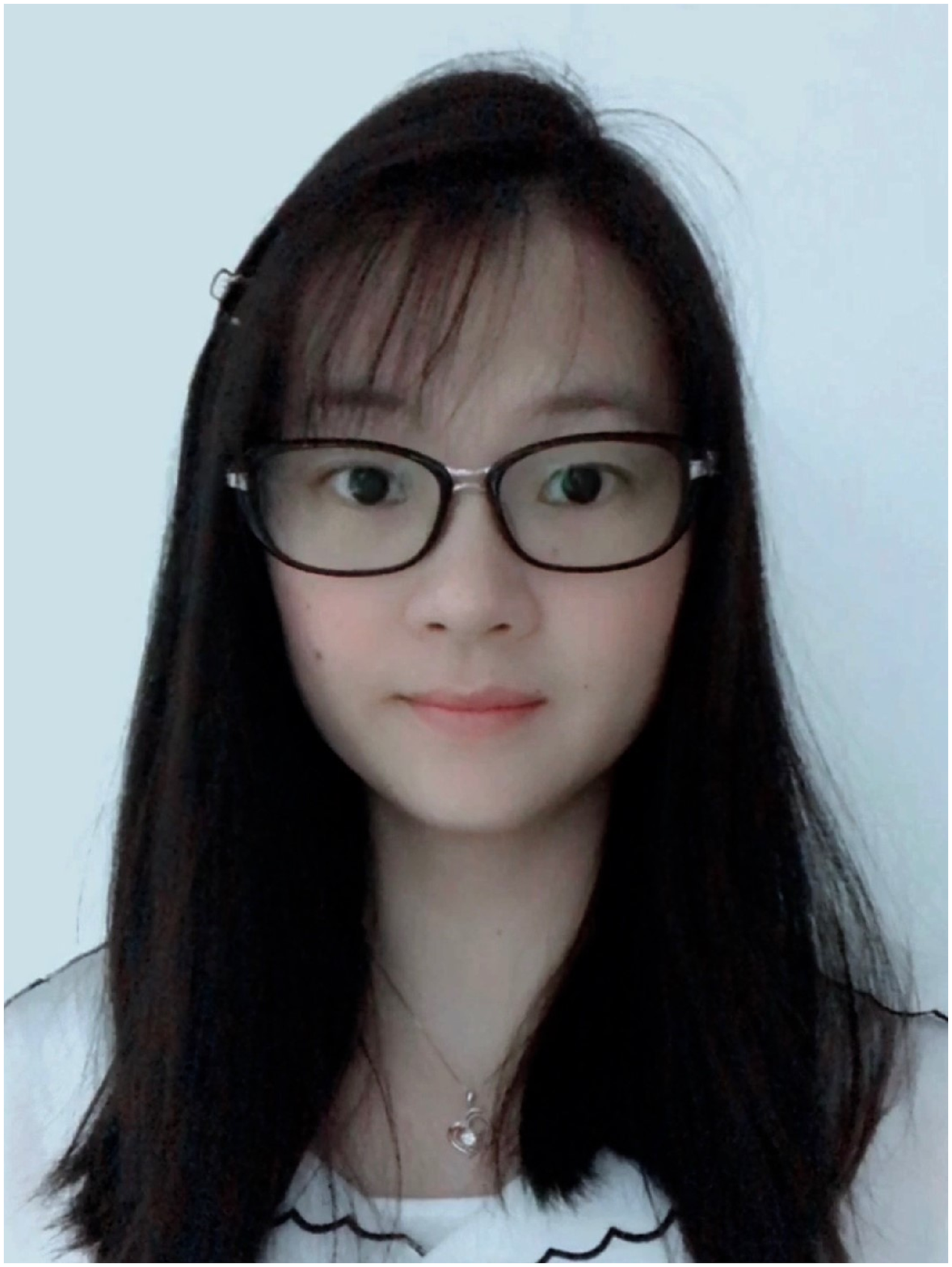}}]{Xuehe Wang} (S'15-M'16)
received her bachelor
degree in mathematics from Sun Yat-sen
University, China in 2011, and her Ph.D. degree in electrical and electronic engineering from Nanyang Technological University, Singapore in 2016. She is an Assistant Professor of Infocomm Technology Cluster with the Singapore Institute of Technology (SIT) since 2019. Before that, she was a postdoctoral research fellow with the Pillar of Engineering Systems and Design, Singapore University of Technology and Design. Her
research interest covers game theory, cooperative control theory, and network economics.
\end{IEEEbiography}

\begin{IEEEbiography}
[{\includegraphics[width=1in,height=1.25in,clip,keepaspectratio]{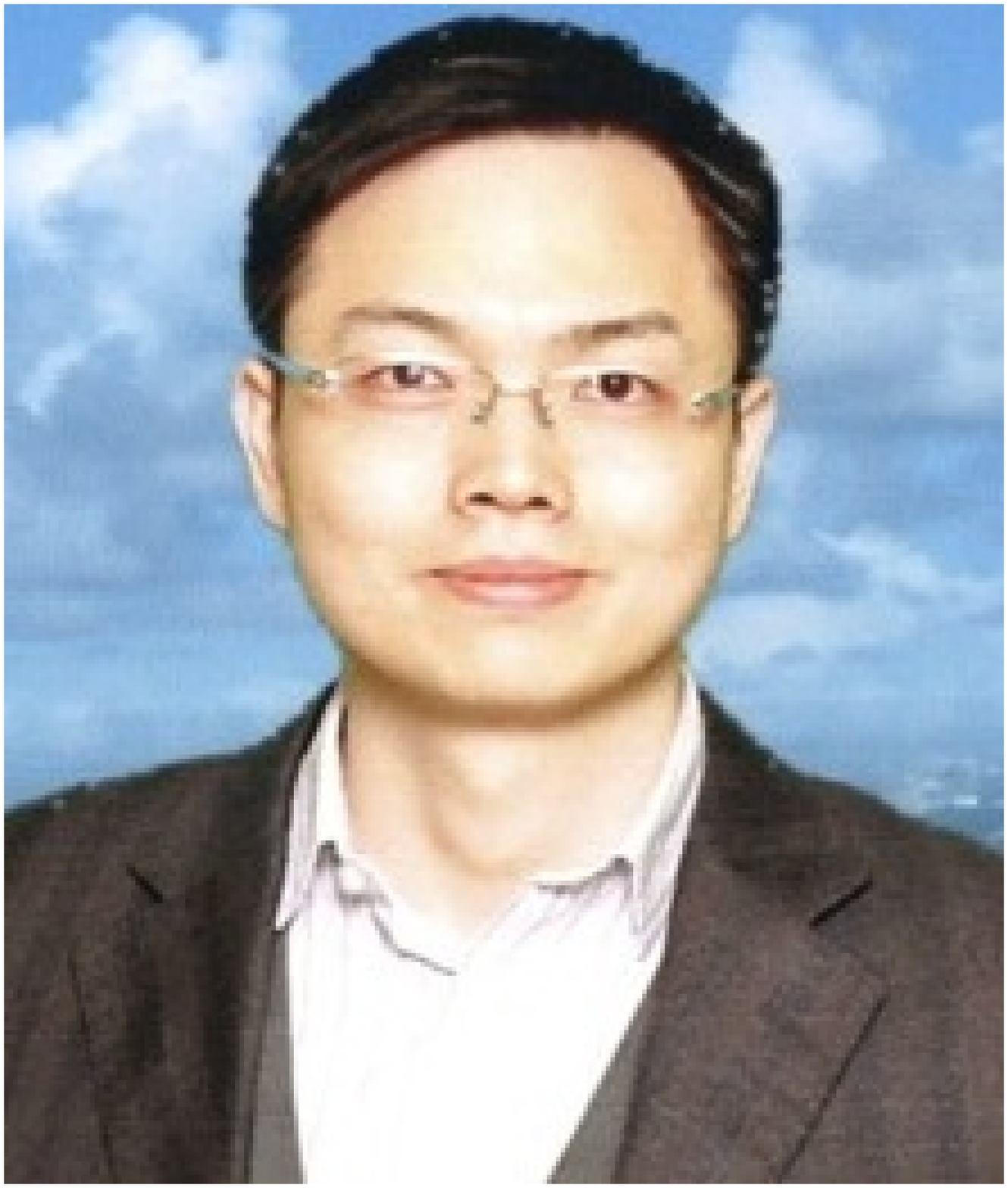}}]{Lingjie Duan} (S'09-M'12-SM'17) received the Ph.D. degree from The Chinese
University of Hong Kong in 2012. He is an Assistant
Professor of Engineering Systems and Design
with the Singapore University of Technology and
Design (SUTD). In 2011, he was a Visiting Scholar at University of California at
Berkeley, Berkeley, CA, USA. His research interests include network economics and game
theory, cognitive communications and cooperative networking, and energy harvesting
wireless communications. He is an Editor of IEEE Transactions on Wireless Communications and IEEE Communications Surveys and Tutorials.
He also served as a Guest Editor of the IEEE Journal on Selected Areas in Communications Special Issue on Human-in-the-Loop Mobile
Networks, as well as IEEE Wireless Communications
Magazine. He received the SUTD Excellence in Research Award in 2016 and the
10th IEEE ComSoc Asia-Pacific Outstanding Young Researcher Award in
2015.
\end{IEEEbiography}

\end{document}